\documentclass[format=acmsmall, review=false, screen=true]{acmart}

\usepackage{booktabs} 
\usepackage{pgf}
\usepackage{colortbl}

\usepackage{subcaption}

\usepackage{color}
\definecolor{red}{cmyk}{0, 0.7808, 0.4429, 0.1412}

\usepackage{inputenc}
\usepackage{multirow}

\newenvironment{packed_item}{
\begin{itemize}
  \setlength{\itemsep}{1pt}
  \setlength{\parskip}{0pt}
  \setlength{\parsep}{0pt}
}{\end{itemize}}

\usepackage[ruled]{algorithm2e} 

\SetAlFnt{\small}
\SetAlCapFnt{\small}
\SetAlCapNameFnt{\small}
\SetAlCapHSkip{0pt}
\IncMargin{-\parindent}

\setcopyright{acmlicensed}
\acmJournal{PACMHCI}
\acmYear{2018} \acmVolume{2} \acmNumber{CSCW} \acmArticle{119} \acmMonth{11} \acmPrice{15.00}\acmDOI{10.1145/3274388}

\setcopyright{acmlicensed}

\acmDOI{10.1145/3274388}

\begin{document}
\title[Endorsements on Social Media]{Endorsements on Social Media: An Empirical Study of Affiliate Marketing Disclosures on YouTube and Pinterest}

\author{Arunesh Mathur}
\affiliation{%
  \institution{Princeton University}
  \streetaddress{35 Olden Street}
  \city{Princeton}
  \state{NJ}
  \postcode{08540}
  \country{USA}}
\email{amathur@cs.princeton.edu}
\author{Arvind Narayanan}
\affiliation{%
  \institution{Princeton University}
  \streetaddress{308 Sherrerd Hall}
  \city{Princeton}
  \state{NJ}
  \postcode{08540}
  \country{USA}}
\email{arvindn@cs.princeton.edu}
\author{Marshini Chetty}
\affiliation{%
  \institution{Princeton University}
  \streetaddress{35 Olden Street}
  \city{Princeton}
  \state{NJ}
  \postcode{08540}
  \country{USA}}
\email{marshini@princeton.edu}

\begin{abstract}
Online advertisements that masquerade as non-advertising content pose numerous risks to users. Such hidden advertisements appear on social media platforms when content creators or ``influencers'' endorse products and brands in their content. While the Federal Trade Commission (FTC) requires content creators to disclose their endorsements in order to prevent deception and harm to users, we do not know whether and how content creators comply with the FTC's guidelines. In this paper, we studied disclosures within affiliate marketing, an endorsement-based advertising strategy used by social media content creators. We examined whether content creators follow the FTC's disclosure guidelines, how they word the disclosures, and whether these disclosures help users identify affiliate marketing content as advertisements. To do so, we first measured the prevalence of and identified the types of disclosures in over 500,000 YouTube videos and 2.1 million Pinterest pins. We then conducted a user study with 1,791 participants to test the efficacy of these disclosures. Our findings reveal that only about 10\% of affiliate marketing content on both platforms contains any disclosures at all. Further, users fail to understand shorter, non-explanatory disclosures. Based on our findings, we make various design and policy suggestions to help improve advertising disclosure practices on social media platforms.
\end{abstract}

%
%

\ccsdesc[500]{Information systems~Sponsored search advertising}
\ccsdesc[500]{Information systems~Social advertising}
\ccsdesc[300]{Information systems~Social networks}
\ccsdesc[500]{Human-centered computing~Empirical studies in HCI}
\ccsdesc[300]{Social and professional topics~Governmental regulations}

%
%

\keywords{Online Advertising; Advertising Transparency; FTC Endorsement Guidelines; Social Media; Sponsored Content}

\maketitle

\renewcommand{\shortauthors}{A. Mathur et al.}

\section{Introduction}
Endorsement-based advertising is one of many advertising strategies that allows Internet content creators---sometimes called \emph{influencers} in marketing discourse---to monetize their content on social media platforms and blogs. Because such advertising often appears in conjunction---and is merged---with content creators' non-advertising content, Internet users encountering these advertisements may not recognize them as such, and may be misled or deceived~\cite{mattress}.

To avoid any such deception and harm to users, several guidelines and regulations around the world---including from Federal Trade Commission (FTC)~\cite{ftc_guidelines} in the United States (US)---require content creators to disclose and describe their relationships with advertisers to users. Having adequate disclosures can better inform users when a piece of content is an advertisement, and ensure that users appropriately weigh up content creators' endorsements. In a recent case, the FTC blocked an endorsement-based advertisement where bloggers were misleading consumers by not disclosing hidden charges for products~\cite{FTC_whats_affiliate_marketing}. The FTC has also warned numerous other content creators---including social media celebrities---who failed to disclose their relationships with brands and companies in their social media posts~\cite{ftc1,ftc2,ftc3,ftc4}.

Unlike traditional online advertising, where disclosures are communicated using standardized icons (e.g., AdChoices~\cite{adchoices} for Online Behavioral Advertising or OBA) or text (e.g., Twitter's \emph{promoted} tweets~\cite{twitterpromoted}) endorsement-based advertising disclosures are relatively unstructured, open-ended in nature, and written primarily by individual content creators. As a result, we know little about how many content creators actually disclose their relationship with advertisers in the first place, and whether users notice and understand the disclosures' underlying message.

In this paper, we examine disclosures accompanying \emph{affiliate marketing}---an endorsement-based advertising strategy that earns content creators money when users click on their customized URLs---on social media platforms. We tackle three primary research questions:
\begin{packed_item}
\item First, how prevalent are disclosures in affiliate marketing content on social media platforms?
\item Second, how do content creators word and frame disclosures in affiliate marketing content?
\item Third, and finally, how effective are these current forms of affiliate marketing disclosures from a user standpoint?
\end{packed_item}

Through our study, we aim to evaluate whether content creators on social media platforms follow affiliate marketing disclosure guidelines set by various regulatory bodies such as the FTC, whether their current disclosures effectively inform users about the underlying endorsement-based advertisement, and if not, determine what steps various stakeholders in the affiliate marketing industry, in policy and in design can take to improve disclosure practices.

To answer our research questions, we first measured the prevalence of affiliate marketing content on two popular social media platforms that serve and enable user-generated content: YouTube and Pinterest. We sampled $\sim$500,000 unique YouTube videos and $\sim$2.1 million unique Pinterest pins, filtered content that contained affiliate URLs, and then examined this content for disclosures. Following this analysis, we conducted a user study with 1,791 Amazon Mechanical Turk (MTurk) users to determine how effective the disclosures we discovered are in practice.

We have four main findings:
\begin{packed_item}
\item First, we find that only $\sim$10\% of all affiliate marketing content on both YouTube and Pinterest contains accompanying disclosures.
\item Second, we find that the affiliate marketing disclosures on these social media platforms can be broadly grouped into three distinct formats, which we term \emph{Affiliate Link} disclosures, \emph{Explanation} disclosures, and \emph{Channel Support} disclosures. We find that \emph{Explanation} disclosures---which the FTC explicitly advocates for using in its guidelines---occur least frequently: only 1.82\% of all affiliate YouTube videos and 2.43\% of all affiliate Pinterest pins contain \emph{Explanation} disclosures.
\item Third, we find that users are able to---by themselves---notice and understand \emph{Explanation} disclosures but not the other types. Disclosures of this type decreased users' perceptions of the content creator favoring the endorsed product by 0.46--0.53 times, and increased users' ability of identifying the underlying advertisement by 2.3--2.8 times. However, we only observed this effect on Pinterest, suggesting that norms and design of social media platforms might influence the efficacy of these disclosures.
\item Fourth, and finally, we find that \emph{Affiliate Link} disclosures are only half as effective as \emph{Explanation} disclosures in communicating the underlying advertisement to users \emph{even when these disclosures are presented to users without them having to look for it}.
\end{packed_item}

We make the following contributions through our study:
\begin{packed_item}
    \item To the best of our knowledge, we are the first to empirically measure the prevalence of affiliate marketing on social media platforms, and the compliance of content creators' accompanying disclosures in affiliate marketing content with the FTC's endorsement guidelines.
    \item We describe a method to detect affiliate marketing content that is broadly applicable to other social media platforms and the Web. As a result of our analysis, we compiled the most comprehensive publicly available list of affiliate marketing companies and their URLs. We make this list available in the Appendix.
    \item We describe a method to retrieve affiliate marketing disclosures from text data. Our method can be easily extended to retrieve disclosures accompanying other forms of endorsement-based marketing techniques such as sponsored content.
    \item We provide new evidence that the wording and framing of disclosures impact user understanding of affiliate marketing advertising on social media platforms.
\end{packed_item}

Based on our findings, we make design and policy suggestions aimed at various stakeholders such as social media platforms and affiliate marketing companies to help enable content creators to easily and clearly disclose advertising relationships. We also outline directions for future work in this area to help users notice and understand affiliate marketing disclosures, and more generally, help enable improved advertising disclosure practices on social media.

\section{Related Work}
In this section, we describe endorsement-based advertising practices including relevant research on advertising disclosures, and place our own study in context.

\subsection{Endorsement-based Advertising on Social Media Platforms}

Native advertising is a form of advertising where advertisements take the form, and structure of non-advertising content, making it harder for users to identify it as such~\cite{hoofnagle2015native,amazeen2018saving,wojdynski2016going,Jiang}. For example, an advertisement might masquerade as a news article along with legitimate news articles on a news website. Endorsement-based advertising on social media platforms is a recent trend within native advertising, where the advertisement is created by a user---termed the content creator---of the platform to generate revenue for themselves rather than the platform itself. Broadly, such advertising manifests in one the following three ways~\cite{wu2016youtube}:

\begin{packed_item}
    \item \emph{Sponsored Content}, in which a content creator partners with an advertiser or merchant to promote a product
    \item \emph{Affiliate Marketing}, in which a content creator posts affiliate URLs to merchants in their content to earn money from the resulting sales
    \item \emph{Product Giveaways}, in which a content creator receives product samples from an advertiser or merchant to promote and review
\end{packed_item}

In this paper, we consider affiliate marketing advertising on social media platforms in depth. We leave the other advertising strategies---\emph{Sponsored Content} and \emph{Product Giveaways}---for future work.

\begin{figure}[t]
    \centering
    \begin{subfigure}[b]{0.3\textwidth}
        \includegraphics[width=\textwidth]{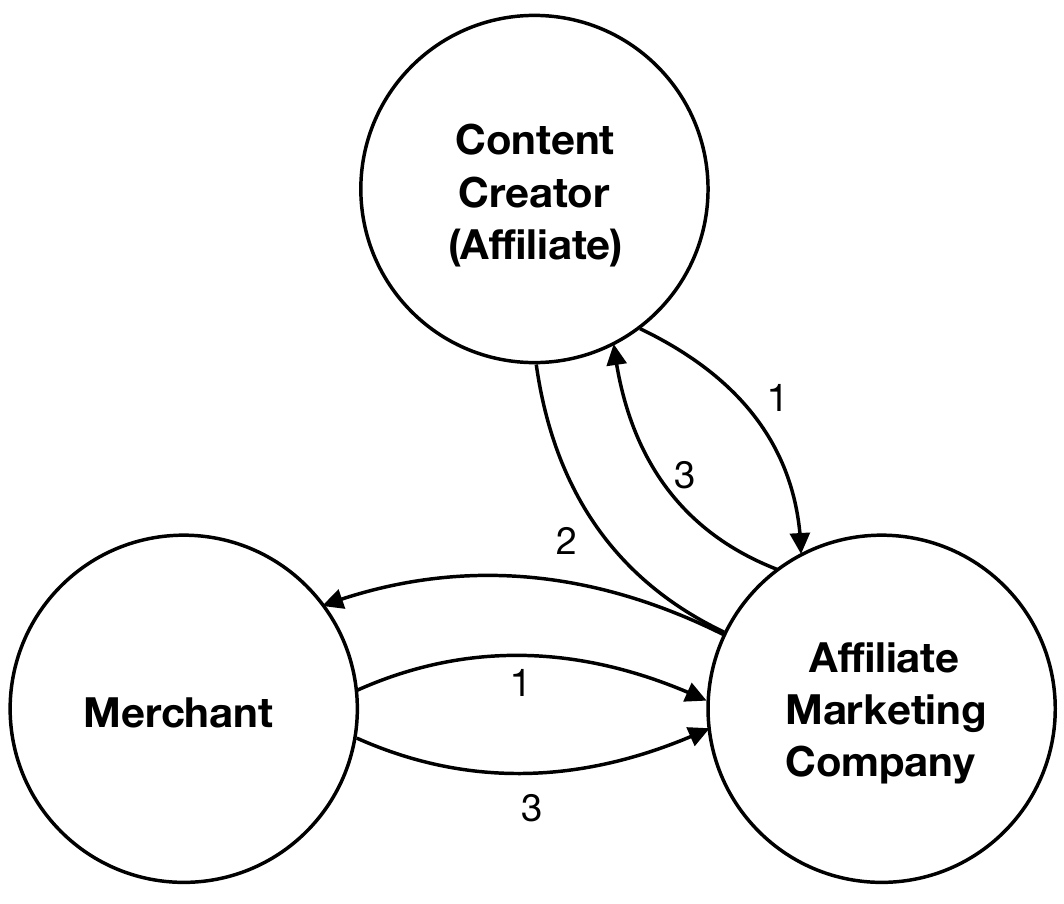}
        \caption{}
        \label{rfidtest_xaxis}
    \end{subfigure}
    \begin{subfigure}[b]{0.34\textwidth}
        \includegraphics[width=\textwidth]{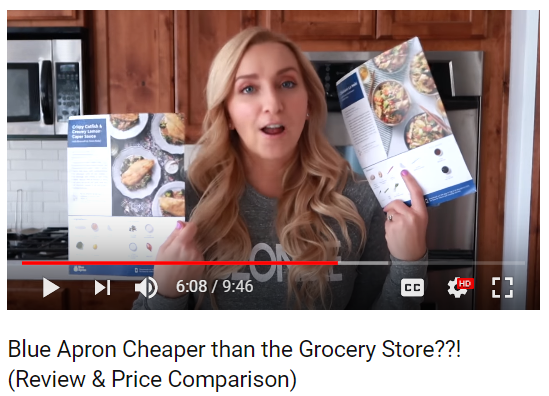}
        \caption{}
        \label{rfidtest_yaxis}
    \end{subfigure}
    \begin{subfigure}[b]{0.34\textwidth}
        \fbox{\includegraphics[width=\textwidth]{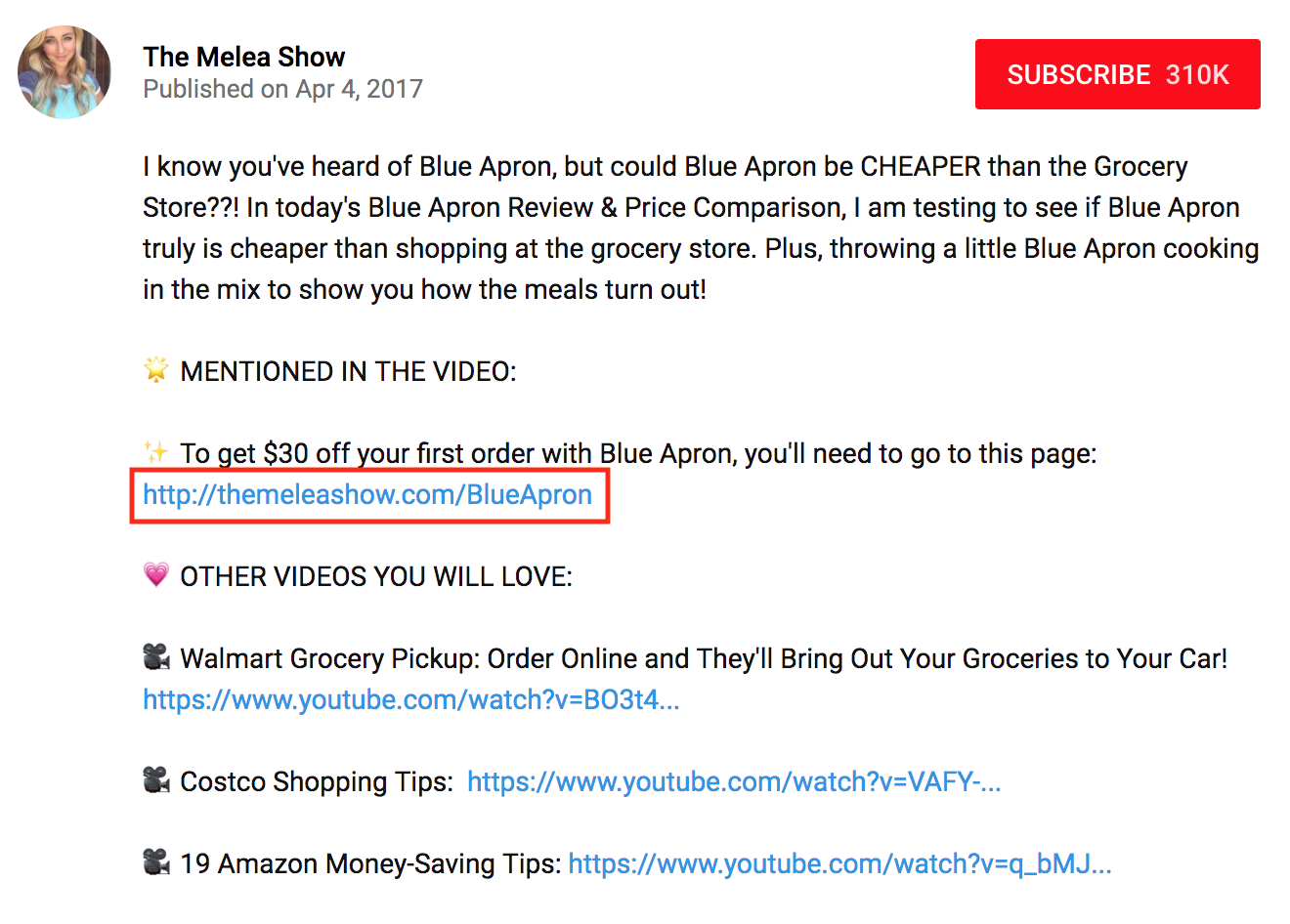}}
        \caption{}
        \label{rfidtest_zaxis}
    \end{subfigure}
    \caption{Overview of Affiliate Marketing and Its Use on YouTube. (a) Affiliate Marketing Ecosystem. (b) YouTube Content Creator Endorsing a Product. (c) Affiliate URL in the Description of the Video (Red Box).}
    \label{fig:aff_mark}
\end{figure}

\textbf{Affiliate Marketing.} Affiliate marketing primarily involves three entities: a content creator, a merchant and an affiliate marketing company. It comprises of two relationships: one between the content creator and the affiliate marketing company, and another between the affiliate marketing company and the merchant. Illustrated in Figure~\ref{rfidtest_xaxis}, affiliate marketing typically works as follows: 

\begin{packed_item}
    \item Merchants and content creators sign up with an affiliate marketing company (1)
    \item Content creators drive sales to the merchant through the affiliate marketing company; the sales are tracked by means of customized URLs---called \emph{affiliate URLs}---that are published by the affiliate marketing company for the content creator to distribute (2)
    \item Each time a sale is made using the affiliate URLs, the merchant pays the content creator a cut of the sale through the affiliate marketing company (3)
\end{packed_item}

Figure~\ref{rfidtest_yaxis} and Figure~\ref{rfidtest_zaxis} illustrate how this works in practice on YouTube. A YouTube content creator endorses a specific brand product (Figure~\ref{rfidtest_yaxis}) in their video, and includes an affiliate URL in the video's description (Figure~\ref{rfidtest_zaxis}). Any purchases through this URL provides a share of the sales---a commission---to the YouTube content creator.

\subsection{Advertising Disclosures}

In the US, the FTC holds advertisers---regardless of the medium where they place their advertisements (e.g., radio, television, print media, or the Internet)---to the \emph{truth-in-advertising} standard, meaning that the advertisements they publish should not be misleading in nature. The FTC requires that if additional information is required to prevent an advertisement from being misleading, then that information must be disclosed to users by adhering to the \emph{clear and conspicuous} standard~\cite{FTC_clear_and_cons}; it investigates any unfair and deceptive practices under Section 5 of the FTC Act~\cite{ftcact}.

The FTC categorizes an advertisement as misleading or deceptive if it conceals its commercial nature from consumers~\cite{FTC_deceptive_ads}. Calling these advertisements---e.g., native advertisements, endorsements, sponsored content---\emph{Deceptively Formatted Advertisements}, the FTC requires that they be disclosed to consumers as advertisements so that consumers can identify them as such. 

In the following sub-sections, we touch upon the landscape of such advertising disclosures.

\subsubsection{Native Advertising Disclosures}

Given that users find it hard to identify and distinguish native advertisements (e.g., search engine advertisements, advertorials) from non-advertising content~\cite{hoofnagle2015native,amazeen2018saving,wojdynski2016going,Jiang}, several studies have examined ways in which the disclosures present on these native advertisements can be designed for better recognition by users. These factors include changing the position, visual appearance and prominence of the disclosure, and the branding of the advertisement content itself. For instance, in a series of studies, Wojdynski \emph{et al.}\ ~\cite{wojdynski2016going,wojdynski2016deceptiveness,wojdynski2017building} found that greater visual prominence of the disclosure such as by increasing text size, color and contrast helped users better recognize the advertisement. Further, even varying the position of the disclosure to the middle and bottom of the article, and using the word \emph{sponsored} was more effective than using a phrase such as \emph{presented by [sponsor]} in helping users recognize the sponsored content.

In a recent report~\cite{ftcstaffreport}, the FTC also discovered that various aesthetic changes to disclosures, such as including borders to surround the advertisement or contrasting the color of the advertisement from other content can significantly increase users' ability to identify native advertisements. Collectively, these studies suggest that effective disclosures need to be visually distinguishable from other content.

\subsubsection{Endorsement-based Advertising Disclosures}

Disclosures requirements for endorsement-based advertising are described in the FTC's endorsement guidelines \cite{ftc_guidelines}. The guidelines state that \emph{if there exists a connection between an endorser and the marketer that consumers would not expect and it would affect how consumers evaluate the endorsement, that connection should be disclosed}. The FTC also requires these disclosures to abide by the \emph{clear and conspicuous} requirement so that consumers can identify them and subsequently decide how much weight to provide to the content creator's endorsement. Further, the disclosures should also not be buried in an \emph{About} or \emph{Terms of Use} page.

For affiliate marketing, the FTC's guidelines state that disclosures need to be placed close to any recommendations and URLs included by the content creator. The guidelines highlight that using \emph{affiliate link} as a disclosure is insufficient, as users may not understand what an affiliate link is. Instead, the FTC recommends using a short phrase such as \emph{I get commissions for purchases made through links in this post} close to the recommendation.

Only recently have studies begun examining how and whether users notice and understand the disclosures made by social media content creators. For instance, Evans \emph{et al.}\ ~\cite{evans} found that stating \emph{paid ad} was more effective than other disclosures in helping users identify advertisements on Instagram. In another study~\cite{van2016effects}, van Reijmersdal \emph{et al.}\ studied how disclosing sponsored content in blogs affected users. They found that after viewing the disclosure, users became more resistant to the endorsement of the blogger.

\subsubsection{Other Types of Advertising Disclosures} Numerous studies have examined advertising disclosure practices in other forms of media such as television content~\cite{structural,grubbs2004adherence,children,boerman2012sponsorship,boerman2014effects}. One set of studies examined differences in television advertisements over time~\cite{grubbs2004adherence} and across demographics~\cite{children}, or content~\cite{structural}. These studies collectively discovered that while the presence of disclosures in advertisements have increased over the years in television advertisements, they still fail to adhere to the \emph{clear and conspicuous} requirement.

Other more recent studies~\cite{boerman2012sponsorship,boerman2014effects} have examined how the length and position of the disclosure affects users' perceptions of sponsored television advertisements. They found that longer disclosures and disclosures made before the sponsored content is presented to users improves users' understanding of the nature of the content. These studies suggest that the timing of when a user sees a disclosure is important for the understanding the meaning of the disclosure.

In OBA---one of the most common forms of online advertising today---users see targeted advertisements based on their interests, demographics, and browsing histories. These online advertisements and the ability to opt-out of them are largely disclosed to users by means of the AdChoices icon---the current advertising industry self-regulation standard. This icon appears in nearly 60\% of online advertisements from just 20 of the top 500 Alexa news websites~\cite{storey2017future}.

Several recent studies~\cite{Leon:2012:OBA:2381966.2381970,ur2012smart,garlach2018m,hastak2010future} have examined what various OBA disclosures communicate to users in practice, collectively revealing several shortcomings of the AdChoices disclosure. Hastak and Culnan~\cite{hastak2010future} were the first to examine this in a survey of nearly 2,600 US adults users. They studied the efficacy of various OBA disclosures in communicating the underlying notice (recognizing the advertisement) and choice (being able to opt-out of tracking) to users. They found that the AdChoices tagline did not work as well other tested taglines in aiding users' comprehension of the disclosure, suggesting that the wording of disclosures is important.

More recently, Ur \emph{et al.}\ ~\cite{ur2012smart} and Leon \emph{et al.}\ ~\cite{Leon:2012:OBA:2381966.2381970} found that non-expert users have several misconceptions about the AdChoices icon, often assuming that clicking on the icon would lead to more information about the advertised product or lead to more advertisements being shown. Garlach and Suthers~\cite{garlach2018m} tested users' understanding of the AdChoices icon on mobile devices and found that participants were unable to locate and click on the icon on the small screens. These studies suggest that users may misunderstand visual indicators for disclosures altogether if they are unfamiliar with what a particular visual is supposed to represent.

\subsection{Summary}
Unlike the above studies, our study is the first to examine affiliate marketing disclosures in the context of endorsement-based advertising on social media. We perform a large scale measurement of social media content that contains affiliate URLs which drive revenue to the creator of the social media content. Further, rather than developing and testing new disclosures, we examine how effective the disclosures content creators currently make are in communicating the underlying endorsement-based advertisement to users. Based on prior work, we test how the placement and wording of disclosures affect user comprehension of disclosures.

\section{Characterizing Affiliate Marketing Disclosures}

\begin{figure}
  \fbox{\includegraphics[width=0.8\textwidth]{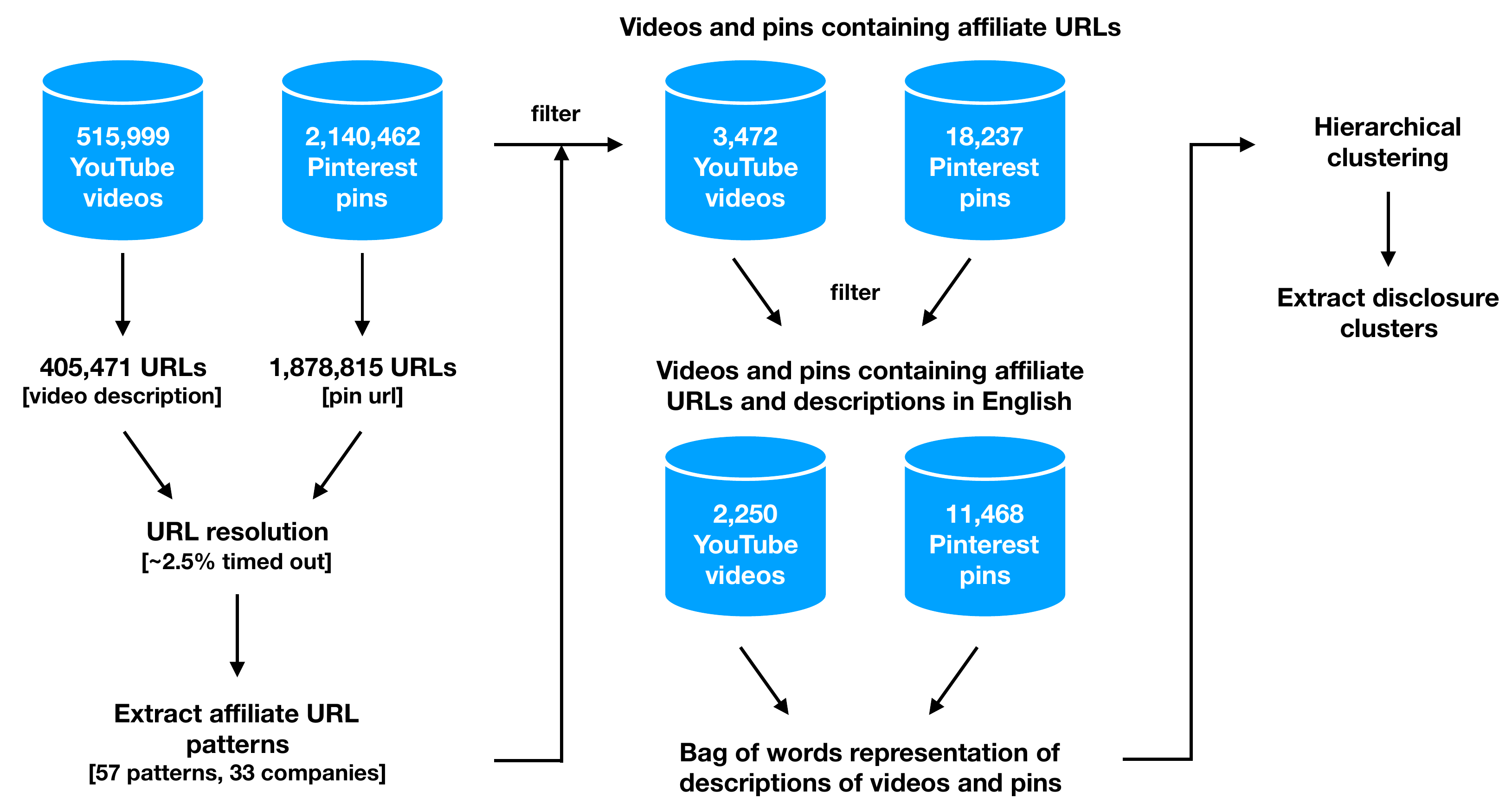}}
\caption{Overview of the Data Collection, Data Cleaning, and Disclosure Extraction Stages.}
  \label{fig:method_collection}
\end{figure}

In this section, we describe how we gathered content from two social media platforms to find affiliate marketing content and disclosures. Through this analysis, we answer our first and second primary research questions: how prevalent are disclosures on affiliate marketing content, and how do content creators disclose their affiliate marketing relationships with merchants to users? Figure~\ref{fig:method_collection} illustrates our data collection and analysis process.

We chose to specifically study YouTube and Pinterest to answer our research questions since these are two popular social media platforms that are designed to share content and reviews. However, our methods are platform agnostic and can be extended to other social media platforms.

\subsection{Method}
\subsubsection{Data Collection}

To answer our research questions, we first gathered data from YouTube and Pinterest. Our goal in this process was to gather the least possibly biased sample of YouTube videos and Pinterest pins in order to accurately model the prevalence of affiliate marketing content and disclosures on these platforms.

We considered three sampling procedures---all of which have been used in past studies---for collecting our data: sampling from large graphs, sampling using keywords, and prefix sampling. Sampling from large graphs involves sampling \emph{Related} content graphs or social network graphs---both of which appear on YouTube and Pinterest as the \emph{Related Videos} and \emph{Related Pins} graph respectively. However, because \emph{Related} content graphs have non-randomly selected edges, which are often biased towards content with high engagement---such as videos with higher view counts in the case of YouTube~\cite{Zhou:2011:CYV:2068816.2068851}---they result in non-uniform samples. Sampling using keywords (e.g., ~\cite{Anthony:2013:AUY:2470654.2466158}) involves gathering samples by querying the search facilities of platforms using a set of compiled keywords. However, samples resulting from this procedure are likely to be biased towards the set of keywords, and in most cases, compiling a set of representative keywords is challenging in itself.

We therefore employed prefix sampling which has previously been used for sampling from YouTube~\cite{Zhou:2011:CYV:2068816.2068851} and Pinterest~\cite{Gilbert:2013:INT:2470654.2481336}. Prefix sampling works by generating---ahead of time---part of the identifier of some record(s) in the population, a \emph{prefix}, which is then used to sample records beginning with that prefix. If the prefixes are uniformly generated, then the resulting samples will be uniform too. This sampling procedure is particularly useful when the search space of identifiers is significantly larger than the number of already issued identifiers, and it is impractical to randomly generate the issued identifiers.

To apply prefix sampling on YouTube, we first observed that YouTube videos are assigned an eleven character long identifier. In order to gather our uniform sample, we first randomly generated video identifiers of length five, a technique used previously~\cite{Zhou:2011:CYV:2068816.2068851}. We then searched for those five character long identifiers using the YouTube Search API~\cite{YouTube_API} which returned a list of videos beginning with that prefix. Similarly on Pinterest, pins are assigned a varying length identifier consisting wholly of numbers. The last five digits of the identifier however, represent a timestamp~\cite{Gilbert:2013:INT:2470654.2481336}. To gather a set of random prefixes, we first retrieved 25 pins---the number of pins returned on average---from the Pinterest \emph{Categories} page~\cite{Pinterest_Categories} for each category. However, these pins were always the most recent pins in that category, and the prefixes they contained were non-random. Therefore, we then fetched all the related pins of these pins and sampled them randomly to create a set of \emph{seed} pins. We then varied the last five digits of these \emph{seed} prefixes from 00000--01500 (arbitrarily) to construct the final sample of pins. In total, we retrieved 515,999 unique YouTube videos and 2,140,462 unique Pinterest pins. While retrieving the videos and pins, we also recorded their characteristics such as what category of content they belonged to, how many times they had been viewed, how many comments had been made on the content, and details about their creators. We gathered this data between August and September 2017.

\subsubsection{Discovering Affiliate Marketing Content}

After gathering the samples of YouTube videos and Pinterest pins, we began the process of identifying affiliate marketing content in the samples. We first compiled a database of all the URLs in the videos and pins, specifically looking for URLs in the description of the YouTube videos and in the Pinterest pins. In total, we compiled a list of 405,471 and 1,878,815 URLs from the videos and pins respectively. To identify affiliate URLs from this set of URLs, we relied on the observation that affiliate URLs contain specific patterns~\cite{chachra2015affiliate}. For instance, Amazon's affiliate URL contains a \emph{tag} parameter which indicates the identifier of the affiliate who stands to gain money from the purchase. However, while Amazon's affiliate URL appears directly on the destination website---the product page on Amazon.com---others may also appear during the intermediate redirects to the destination. Further, unlike Amazon, affiliate URL patterns may not necessarily only emerge as URL parameters; they may also be present in other parts of the URL including its path and sub-domain.

To ensure that our analysis accounted for all such cases, we first resolved all the URLs in our database, following both server-side and client-side redirects (HTTP 3XX, Meta refresh), and recorded the resulting intermediate URLs and all HTTP response codes. A total of $\sim$2.5\% of the URLs across both datasets failed to resolve either due to timeouts or HTTP 404 error codes; we ignored these URLs. Next, we performed a frequency analysis using each resolved URL's domain, sub-domain, path, and parameters, creating a list of commonly occurring patterns sorted by decreasing order of appearance (counts). We reasoned that if there existed any patterns across the URLs we resolved and visited, our frequency analysis would capture and bubble these to the top of the list. Starting with the sub-domain and path, we first recorded how many sub-domains (paths) each domain appeared with. A high number of sub-domains (paths) would signal that an affiliate marketing company likely caters to different merchants through unique sub-domains (paths), or that a constant sub-domain (path) appears as part of the affiliate URL. Next, we turned our attention to domains and their URL parameters. Rather than recording the number of parameters associated with each domain, we recorded the number of times a domain appeared with a URL parameter. A frequent co-occurrence of domains and URL parameters would signal that the parameter conveys some information about the content creator to the affiliate marketing company. 

Because these lists contained a high number of false positives, we manually scanned each list, examining which of the domains, sub-domains, paths, parameters corresponded to affiliate marketing companies. To aid our examination, we used a combination of Google Search results, the WHOIS database, and the FMTC affiliate database~\cite{FMTC}, which provides a mapping from merchants to affiliate programs. Where possible, we also signed up on these programs as content creators to validate our findings about the affiliate programs. To limit the effort required to manually examine these lists, we only examined those combinations of domains and sub-domains/paths/parameters that appeared at least 15 times.

\subsubsection{Discovering Affiliate Marketing Disclosures}
After finalizing the affiliate URL patterns, we first filtered the list of resolved URLs to only retain those corresponding to these patterns. We then filtered the YouTube videos and Pinterest pins datasets to only those containing the affiliate URLs. Following this step, we began the process of extracting the disclosures from these videos and pins. We could have expected to find the disclosures---in theory---either during the course of the videos, the image of the pins, or the in the description boxes surrounding the videos and pins.

As a first step, we randomly sampled 20 videos and pins each from our filtered dataset of affiliate marketing videos and pins. We then examined these videos and pins to look for any affiliate marketing disclosures during the course of the videos and the pins' images. Because we found no disclosures in these random samples, we resorted to extracting disclosures in the descriptions of the videos and pins.

To extract the disclosures present in these videos' and pins' descriptions, we first split each description by its newlines and then by the sentences contained in each newline. We then tokenized each resulting sentence into a bag-of-words representation, and clustered the sentences using hierarchical clustering~\cite{Rokach2005} with the euclidean distance metric. We chose a fairly low cut-off for the clusters based on the idea that the relevant smaller clusters containing the affiliate disclosures may already have been formed at that cut-off. We then manually examined these clusters one after the other, and recorded ones that contained disclosures pertaining to affiliate marketing. For this analysis, we only considered those descriptions that were written in English; doing so retained 64.8\% and 62.8\% of all affiliate videos and pins respectively.

\subsubsection{Limitations}

Our analyses of affiliate marketing content and disclosures has limitations. First, our method to discover affiliate URLs should be considered a lower bound on the number of affiliate URLs and consequently on affiliate marketing companies since we ignored those URL co-occurrences that appeared less than 15 times. Though we may have missed less prevalent companies owing to our frequency analysis, we are confident that our analysis presents a close approximation of affiliate URL prevalence. We also did not consider those affiliate programs that use coupon codes to track and attribute sales. Second, our method to discover affiliate marketing disclosures was limited to descriptions written in English. As a consequence, our findings may not generalize to other languages. In our data set, the descriptions of $\sim$40\% of all affiliate marketing videos and pins were written in a non-English language.

\subsection{Findings}
In the following sections, we present our findings, describing the characteristics of affiliate marketing content on both platforms, the types of disclosures made by content creators, and the characteristics of these disclosures.

\subsubsection{Examining the Affiliate Marketing Landscape}
Through our analysis, we compiled a total of 57 unique affiliate URL patterns from 33 different affiliate marketing companies---the most comprehensive publicly available list of its kind. This list facilitated automated detection of affiliate marketing content on both the social media platforms in our work and can be used to detect similar content on other platforms in future studies. We found at least one affiliate URL in a total of 0.67\% or 3,472 videos on YouTube, and a total of 0.85\% or 18,237 pins on Pinterest.

Table~\ref{tab:affiliate} in the Appendix lists the affiliate marketing companies we discovered along with their URL patterns, and the number of times we observed their presence across all URLs resolutions in our entire dataset. Note that a YouTube video may contain multiple affiliate URLs in its description unlike a Pinterest pin, which only contains one URL (that is, the pinned URL).

Across both YouTube and Pinterest, Amazon's Associate Program\footnote{https://affiliate-program.amazon.com/} had the largest presence (YouTube = 7,308, Pinterest = 7,368), closely followed by AliExpress' Affiliate Program\footnote{https://portals.aliexpress.com} (YouTube = 2,167, Pinterest = 785). In addition to Amazon, we discovered other merchants that hosted in-house affiliate programs, as opposed to explicitly redirecting through an affiliate marketing company. For instance, Booking.com\footnote{https://www.booking.com/affiliate-program/v2/index.html} and Apple\footnote{https://www.apple.com/itunes/affiliates/} marketed products through their own affiliate programs. Some of the companies in our list are easily recognizable but many are not likely to be familiar to users.

\begin{figure}[t]
    \centering
    \includegraphics[width=\textwidth]{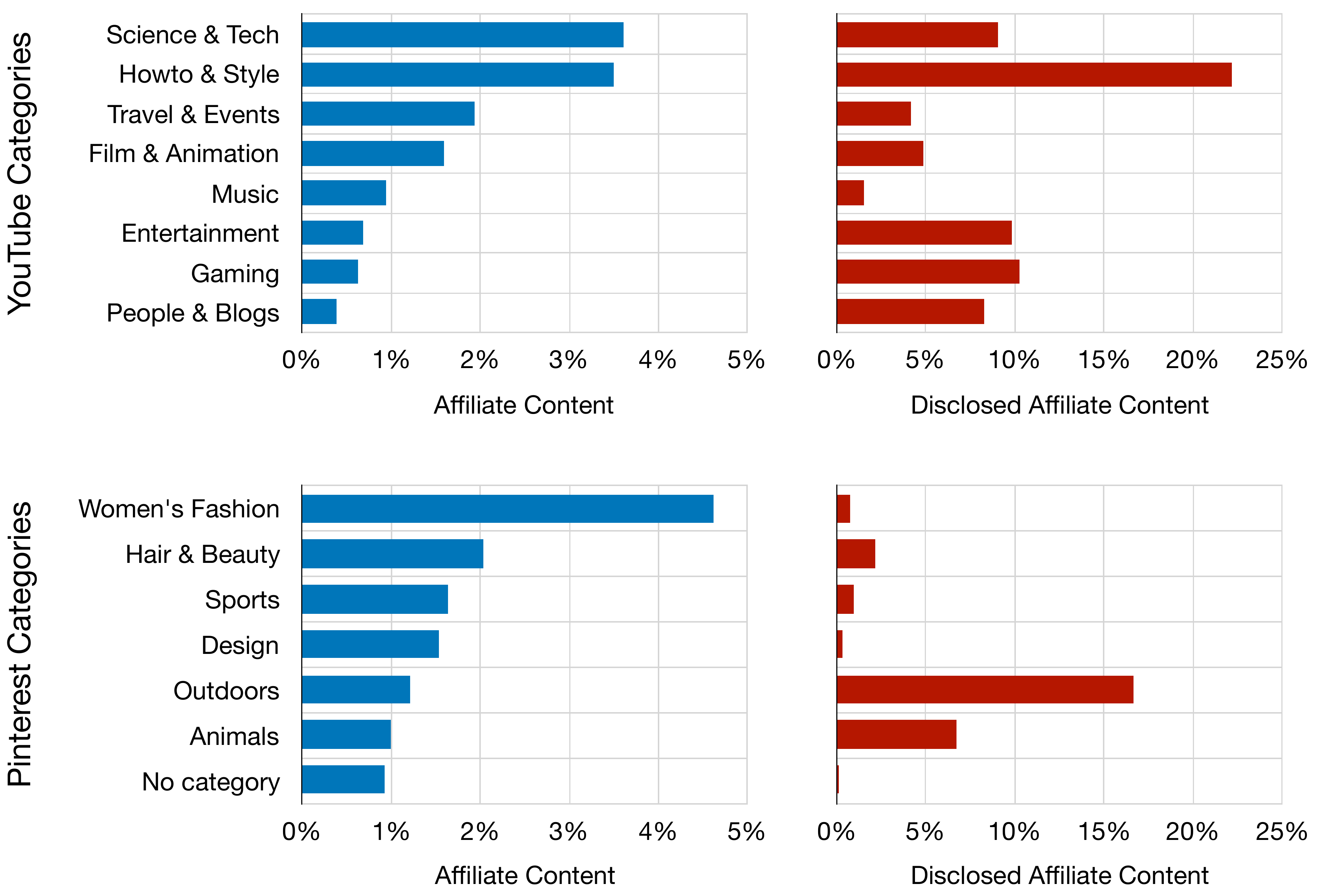}
    \caption{Percentage of Affiliate Marketing Content Across All Content (Blue Bar Charts) and Percentage of Disclosed Affiliate Content (Red Bar Charts) Within Each YouTube and Pinterest Category. Only Categories With More Than 100 Affiliate Marketing Videos and Pins Are Shown.}
    \label{fig:gra}
\end{figure}

\subsubsection{Fashion, Beauty, and Technology Content Contain Highest Affiliate Marketing}
\label{sec:content}

Next, we analyzed how much content within each category---as defined by the platform---on YouTube and Pinterest contained affiliate URLs, and ranked the resulting categories. For this analysis, we only ranked those categories that had at least 100 affiliate marketing videos or pins.

The blue bar charts in Figure~\ref{fig:gra} illustrate our findings. Across both platforms, we found that affiliate marketing was most commonly found across content within technology, fashion, and beauty related categories. Prevalence of affiliate marketing across YouTube's \emph{Science \& Technology} and \emph{Howto \& Style} categories stood at 3.61\% and 3.49\% respectively. Similarly, prevalence of affiliate marketing across Pinterest's \emph{Women's Fashion} and \emph{Hair \& Beauty} categories stood at 4.62\% and 2.04\% respectively.

We suspect the presence of affiliate marketing in these categories may be attributed to at least two different reasons. First, technology products, apparel, and beauty/cosmetics items are some of the most common products consumers shop for online~\cite{productcat1,productcat2,productcat3}, so their presence in affiliate marketing might be indicative of this larger trend. Second, recent surveys have shown that users proactively seek out reviews before shopping for technology products online, and therefore this may be reflected in affiliate marketing content on these platforms~\cite{productcat4}.

\subsubsection{Affiliate Marketing Content Has Higher User Engagement}

Next, we examined how affiliate and non-affiliate marketing content correlated with user engagement metrics such as view, like, and comment counts. We conducted Mann--Whitney \emph{U} tests to assess statistical significance. Accounting for multiple testing using Bonferroni correction, we tested for significance at the 0.01 level.

Across both YouTube and Pinterest, we noted a common thread: videos and pins with affiliate marketing content correlated with higher user engagement metrics. That is, these videos and pins had more comments, were liked more, and viewed more often than other non-affiliate marketing content. On YouTube, affiliate marketing videos correlated with longer duration length ($U \sim 7.95 \times 10^8$, $p < 0.0001$), higher view counts ($U \sim 7.72 \times 10^8$, $p < 0.0001$), higher like counts ($U \sim 6.96 \times 10^8$, $p < 0.0001$), and higher dislike counts ($U \sim 6.52 \times 10^8$, $p < 0.0001$). Similarly, on Pinterest, affiliate marketing pins correlated with higher repin counts ($U \sim 1.93 \times 10^{10}$, $p < 0.0001$). We could not directly compare the like and comment counts on Pinterest since the Pinterest API returned both as zero for all the pins in our dataset. Given the higher user engagement, affiliate marketing content is likely to surface through recommendations algorithms, and ensuring content creators disclose these endorsements becomes an even more pressing issue.

\begin{table}[t]
\centering
\caption{Prevalence of Various Affiliate Disclosure Types On YouTube and Pinterest. Prevalence is Computed Across All Affiliate Content.}
\label{tab:prevdisc}
\begin{tabular}{@{}lccl@{}}
\toprule
\textbf{Disclosure} & \textbf{Platform} & \textbf{Prevalence (\%)} & \multicolumn{1}{c}{\textbf{Example}}                                                                                                                                \\ \midrule
\multirow{2}{*}{Affiliate Link}                                    & YouTube           & 7.02                     & Affiliate links may be present above                                                                                                                                \\
                                                                   & Pinterest         & 4.60                     & (aff link)                                                                                                                                                          \\
                                                                   \cline{2-4} 
\multirow{2}{*}{Explanation}                                       & YouTube           & 1.82                     & \begin{tabular}[c]{@{}l@{}}This video contains affiliate links, which means\\ that if you click on one of the links, I'll\\ receive a small commission\end{tabular} \\
                                                                   & Pinterest         & 2.43                     & \begin{tabular}[c]{@{}l@{}}(This is an affiliate link and I receive a commission\\ for the sales)\end{tabular}                                                      \\
                                                                   \cline{2-4} 
\begin{tabular}[c]{@{}c@{}}Channel \\ Support\end{tabular}         & YouTube           & 2.44                     & \begin{tabular}[c]{@{}l@{}}AMAZON LINK: (Bookmark this link to support\\ the show for free!!!)\end{tabular}                                                         \\ \bottomrule
\end{tabular}
\end{table}

\subsubsection{Three Types of Affiliate Marketing Disclosures}

Across YouTube and Pinterest, we discovered that 10.49\% and 7.03\% of affiliate marketing videos and pins respectively were disclosing the affiliate URLs in their content descriptions as such to users. We found three distinct types of disclosure clusters which we describe below; Table~\ref{tab:prevdisc} summarizes our findings. 

\textbf{\emph{Affiliate Link} Disclosures}: The first type of disclosures communicated to users that affiliate URLs were present in the content. On YouTube, these disclosures appeared in the video description either as blanket disclosures---a single disclosure across the entire description---or as disclosures highlighting the individual affiliate URLs, or both. Exactly 7.02\% of affiliate marketing videos were disclosing in this manner. The following statements describe how YouTube content creators made these types of disclosures:

\begin{itemize}
\item \emph{Affiliate links may be present above}
\item \emph{Some of the links may be affiliate links}
\item \emph{(Disclosure: These are affiliate links)}
\item \emph{*Amazon link(s) are affiliate links}
\end{itemize}

On Pinterest, \emph{Affiliate Link} disclosures appeared in the description of 4.60\% of all affiliate marketing pins. Unlike YouTube, content creators' disclosures did not point to specific URLs, since the Pinterest pins only contained the pinned URL. The following statements describe how Pinterest content creators made these types of disclosures:

\begin{itemize}
    \item \emph{(aff link)}
    \item \emph{(affiliate)}
    \item \emph{\#affiliatelink}
    \item \emph{This is an Amazon Affiliate link}
\end{itemize}

\textbf{\emph{Explanation} Disclosures}: The second type of disclosures content creators made offered users a verbose explanation about affiliate marketing and affiliate URLs, and described how clicking on the URLs impacts users viewing the content. In comparison to \emph{Affiliate Link} disclosures, content creators who used \emph{Explanation} disclosures included more details, and often quoted specific merchants or affiliate marketing companies. On YouTube, these disclosures appeared in the descriptions of 1.82\% all affiliate marketing videos. The following statements describe how YouTube content creators made these types of disclosures:

\begin{itemize}
    \item \emph{This video contains affiliate links, which means that if you click on one of the links, I'll receive a small commission}
    \item \emph{I am an affiliate with eBay, Amazon, B\&H and Adorama, which means I get a small commission when you buy through my links}
    \item \emph{**Links that start with http://rstyle, Beautylish \& MUG links are affiliate links, I do earn a small commission when you purchase through them, which helps me purchase products for review \& improve my channel}
\end{itemize}

On Pinterest, \emph{Explanation} disclosures appeared in the descriptions of 2.43\% of all affiliate marketing pins. The following statements describe how Pinterest content creators made these disclosures:

\begin{itemize}
    \item \emph{(This is an affiliate link and I receive a commission for the sales)}
\end{itemize}

\textbf{\emph{Channel Support} Disclosures}: The third type of disclosures content creators made communicated to users that they would be supporting the channel by clicking on the affiliate URLs, without clearly explaining how. Only content creators on YouTube made these type of disclosures, which appeared in the descriptions of 2.44\% of all affiliate marketing videos. The following statements describe how YouTube content creators made these types of disclosures:

\begin{itemize}
    \item \emph{AMAZON LINK: (Bookmark this link to support the show for free!!!)}
    \item \emph{Support HWC while shopping at NCIX and Amazon}
    \item \emph{Purchase RP here and help support this channel via the amazon affiliate program}
    \item \emph{Shop using these links to support the channel}
\end{itemize}

To summarize, we found that only $\sim$10\% and $\sim$7\% of videos and pins respectively with affiliate marketing URLs contained disclosures by content creators. When present, these disclosures were of three types: \emph{Affiliate Link} disclosures, \emph{Explanation} disclosures, and \emph{Channel Support} disclosures. Our results show that \emph{Explanation} disclosures, which the FTC explicitly advocates content creators to use in its endorsement guidelines---and are also the most effective as we show our in user study---appeared least frequently.

\subsubsection{Affiliate Marketing Disclosure Prevalence by Content Category}

We also examined how the prevalence of content creators' affiliate marketing disclosures varied across the videos' and pins' categories (as we described in Section~\ref{sec:content}). To account for the number of content creators contributing towards the disclosures in each category, we scaled the overall disclosure percentage by the ratio of number of unique content creators accounting for those disclosures and the number of videos and pins containing disclosures. If the all the videos and pins containing disclosures were accounted for by unique creators, then the percentage remained unchanged.

The red bar charts in Figure~\ref{fig:gra} illustrate our findings. We observed some variance in the percentage of affiliate marketing content on both YouTube and Pinterest that contained any form of disclosure, with some outliers on both platforms. At about 22.5\%, affiliate marketing videos in the \emph{Howto \& Style} category on YouTube contained more disclosures than affiliate marketing videos in the other categories. In stark contrast, disclosure rates in the corresponding \emph{Hair \& Beauty} and \emph{Women's Fashion} categories on Pinterest was relatively low (0.5\%--2.5\%). Similarly, affiliate marketing pins in the \emph{Outdoors} category on Pinterest had an unusually high overall disclosure percentage, at about 15\%. We suspect the relatively high rate of disclosures on YouTube's fashion category may partially be explained by recent awareness of the FTC's guidelines among fashion and beauty vloggers on YouTube~\cite{ftc1,ftc5}. However, as stated before, our analysis only examined disclosures written in English; the distribution of disclosures in affiliate content for other languages remains an area for future exploration.

\section{User Study of Affiliate Marketing Disclosures}
Having discovered the different types of affiliate marketing disclosures on YouTube and Pinterest, we turned to our third primary research question: How well do users notice and understand current forms of these affiliate marketing disclosures? More specifically, we asked:

\begin{packed_item}
    \item Do users notice the disclosures present in the description of the affiliate marketing content?
    \item Does the position of the disclosure affect whether users notice the disclosure?
    \item Regardless of whether users notice the disclosures, what do these disclosures communicate to users?
    \item How much do users think of the product embedded in affiliate marketing content (e.g., Figure~\ref{rfidtest_yaxis}) as an endorsement by the content creator?
    \item How does the presence of a disclosure change users' perception of the endorsement?
\end{packed_item}

To answer these questions, we conducted online experiments on the MTurk platform. We describe the logistics of the experiment, our subsequent data analysis, and findings in the following sections. 

\subsection{Method}

\subsubsection{Experiment Instrument}

\begin{figure}
  \includegraphics[width=\textwidth]{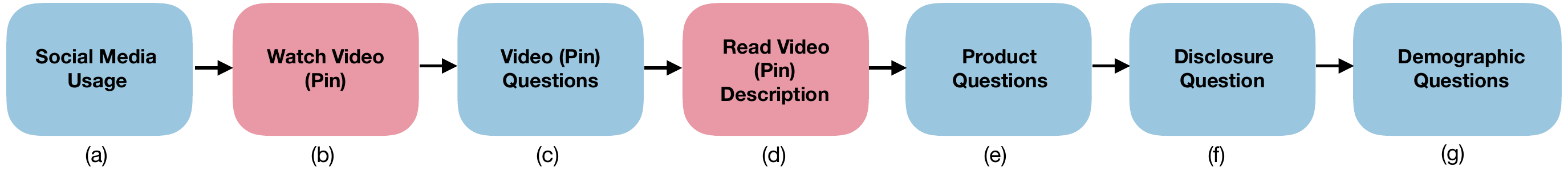}
\caption{The Overall Structure and Flow of the YouTube and Pinterest Experiments. Blue Boxes Represent Sections of the Experiment Where Participants Only Answered Questions; Red Boxes Represent Sections of the Experiment Where Participants Only Viewed Content.}
  \label{fig:experiment}
\end{figure}

We conducted two online randomized controlled experiments on MTurk---one each for YouTube and Pinterest---employing a between-subjects full-factorial design in each experiment. We examined the effect of two independent variables---\emph{disclosure type} and \emph{video/pin}---on several dependent variables, which we describe in the following paragraphs. The first independent variable, \emph{disclosure type} was a fixed factor and its levels consisted of the disclosure types that we found in the previous analysis. We further varied the \emph{disclosure type} by position to appear either at the top or the bottom of the content description. In the YouTube experiment, this resulted in six levels with three disclosure types---(\emph{Affiliate Link}, \emph{Explanation}, and \emph{Channel Support})---spread across two disclosure positions (top, bottom). We chose the following disclosure statements for each disclosure type:

\begin{packed_item}
\item Control: [No Disclosure]
\item Affiliate Link Disclosure: \emph{Affiliate links may be present}
\item Explanation Disclosure: \emph{This video contains affiliate links, which means that if you click on one of the links, I receive a commission for the sales}
\item Channel Support Disclosure: \emph{Shop using these links to support the channel}
\end{packed_item}

Similarly, in the Pinterest experiment, the \emph{disclosure type} independent variable contained four levels with two disclosure types---(\emph{Affiliate Link} and \emph{Explanation})---spread across two disclosure positions (top, bottom). We chose the following disclosure statements for each disclosure type (Recall, Pinterest had no \emph{Channel Support} type disclosures):

\begin{packed_item}
\item Control: [No Disclosure]
\item Affiliate Link Disclosure: \emph{Affiliate link}
\item Explanation Disclosure: \emph{This is an affiliate link and I receive a commission for the sales}
\end{packed_item}

Our choice of the disclosure statements warrants a justification. Our goal with the user study was to determine the efficacy of the types of disclosures currently being made; we did not set out to design and test new hypothetical affiliate marketing disclosures, and we leave doing so for future work. This reasoning is reflected in our choices above: for each level in the \emph{disclosure type} factor---whether in the YouTube or Pinterest experiment---we picked an instance of the disclosure that we thought would best represent disclosures in that category on that platform, but at the same time would generalize across videos and across pins.

The second independent variable \emph{video/pin}, reflected the videos and pins participants watched during the experiments. This independent variable was a random factor in that we created its levels from the affiliate marketing videos and pins in our dataset. We considered \emph{video/pin} as a random factor since we wished to examine how our dependent variables varied considering the differences across videos and pins, rather than examining the effect of video/pin on the dependent variables per se. To construct the levels of the \emph{video/pin} random factor we sampled five videos and five pins from the list of affiliate marketing videos and pins our dataset. However, in order to limit the duration of the YouTube experiment, we only sampled the five videos from around the overall median affiliate marketing video length ($\sim$210 seconds); we added no such constraints to the five randomly selected affiliate marketing pins in the Pinterest experiment. We specifically chose five videos and pins to balance experimental overhead; a larger number would require more participants for a between-subjects design. None of these videos and pins contained any disclosures during the video or on the pin's image.

With these two independent variables and including the control, we arrived at a (5 X 7) and a (5 X 5) full-factorial design in the YouTube and Pinterest experiments respectively. In each experiment, we randomly assigned participants to one of the \emph{cells} of the experiments.

\begin{figure}[t]
    \centering
    \begin{subfigure}[b]{0.45\textwidth}
        \fbox{\includegraphics[width=\textwidth]{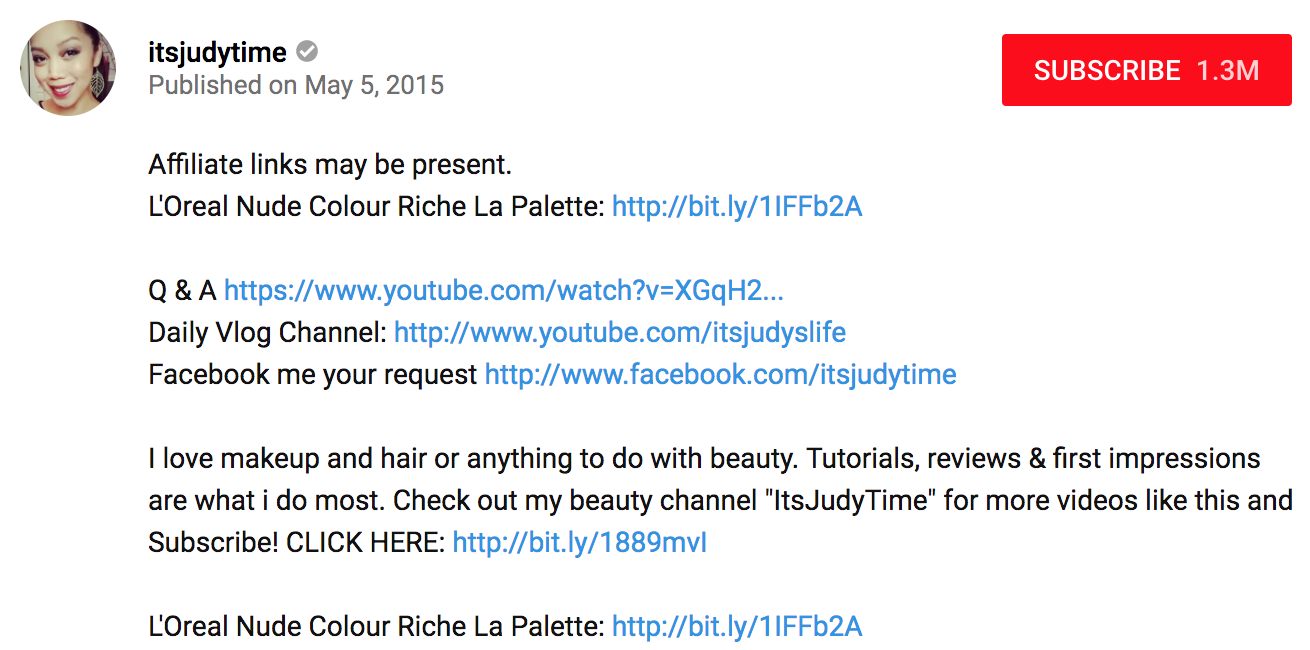}}
        \caption{\emph{Affiliate Link} [Top] Condition (cropped)}
        \label{t1}
    \end{subfigure}
    \hspace{3mm}
    \begin{subfigure}[b]{0.45\textwidth}
        \fbox{\includegraphics[width=\textwidth]{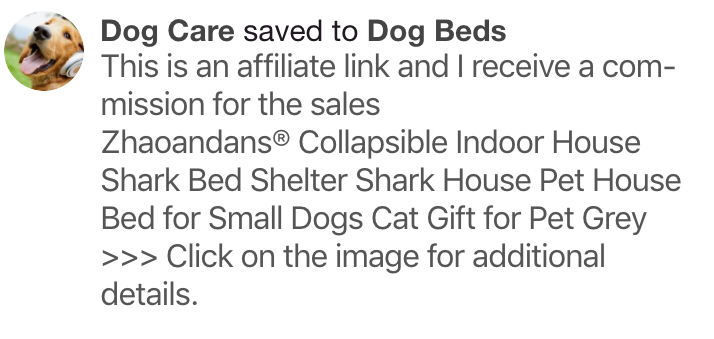}}
        \caption{\emph{Explanation} [Top] Condition}
        \label{t2}
    \end{subfigure}
    \caption{Examples of the Treatment Conditions in the (a) YouTube Experiment and (b) Pinterest Experiment. The Disclosure Statements Were Added to the Top of the Descriptions in these Treatment Conditions.}
    \label{fig:treat}
\end{figure}

Figure~\ref{fig:experiment} illustrates the overall structure of the YouTube and Pinterest experiments. Each experiment contained five parts. 

\textbf{Part One.} In the first part of the experiment (Figure ~\ref{fig:experiment}a), we gathered details about participants' social media use. Specifically, we asked participants whether they had an account on a list of popular social media platforms including YouTube and Pinterest, how often they visited those platforms, and how often they posted content on those platforms. 

\textbf{Part Two.} In the second part of the experiment, participants either watched a YouTube video or looked at a Pinterest pin (Figure ~\ref{fig:experiment}b) which was embedded in the experiment page without its textual description. Immediately after this step, we asked participants' opinions about the content they watched (\emph{video, pin impression}) on a Likert scale (extremely negative--extremely positive; 5 point), and by means of an open-ended question (Figure ~\ref{fig:experiment}c). Based on how accurately they described the content, the responses to this question helped us ascertain whether participants actually viewed the content. In the YouTube experiment, we also logged participants' interactions with the embedded video player (play, pause, stop) to determine whether they watched the video.

\textbf{Part Three.} In the third part of the experiment (Figure ~\ref{fig:experiment}d) we evaluated users' perceptions of the product embedded in the video or pin as an advertising endorsement by the content creator. In this part, we presented participants with a screenshot of the description of the video or pin they had just viewed. Depending on the condition, this was---including the control---either one of the seven conditions in YouTube experiment or one of the five conditions in the Pinterest experiment. Figure~\ref{fig:treat} shows one of treatments conditions for one video and one pin from both the YouTube and Pinterest experiments.

In the following page (Figure ~\ref{fig:experiment}e), we presented participants with an image of the merchant's website that contained the product that was linked using an affiliate URL in the video or pin's description. We then asked participants to report their impression of the product (\emph{product impression}) on a Likert scale (extremely negative--extremely positive; 5 point), and how much they thought the content creator of the video/pin favored the product (\emph{content creator favors}) on a Likert scale (does not favor at all--strongly favors; 5 point). We used the latter as a measure of users' perceptions of the underlying endorsement by content creators.

To understand whether users notice the disclosures present in the descriptions of affiliate marketing content, we then asked participants how likely they thought there was a relationship between the content creator and the organization selling the product (\emph{content creator relationship}) on a Likert scale (extremely unlikely--extremely likely; 5 point). We also included an open-ended question asking participants to explain their reasoning behind their answer to the previous question.

\textbf{Part Four.} In the fourth part of the experiment (Figure ~\ref{fig:experiment}f), we presented the disclosure statement---which again varied depending on the treatment condition---to participants and asked them to explain what the statement meant in their own words by means of an open-ended question (\emph{explain}). Using this question, we evaluated what these disclosures communicate to participants when they are presented---and are paying attention---to the disclosure statement.

\textbf{Part Five.} Finally, the fifth part of the experiment (Figure ~\ref{fig:experiment}g) contained demographic questions.

Throughout the experiment(s), we referred to the product that was linked using the affiliate URL and the merchant as an \emph{item} and \emph{organization selling the item} to ensure that our questions appeared neutral and not leading. 


\subsubsection{Experiment Pilot and Deployment}

We piloted the experiment(s) with 10 participants on UserBob\footnote{http://userbob.com/}, a usability testing website that employs MTurk workers for think-aloud screen-capture sessions. During the sessions, we asked participants to describe---in their own words---what the purpose of each question was, and whether they experienced any difficulty in answering them. Using this feedback, we restructured and refined the questions.

We then deployed the experiment(s) on MTurk as a \emph{Give us your opinion about this YouTube Video (or Pinterest Pin)} task. We used this neutral phrase to avoid leaking any information about our motives behind the experiment. We required that participants completing the experiment be from the US, and have a MTurk approval rating of 95\% or higher. Accounting for the minimum federal wage in the US (\$7.25/hour), we paid participants \$1.25 for completing the YouTube experiment and \$0.75 for completing the Pinterest experiment since these took no longer than 10 and 5 minutes to complete respectively. We gathered all the data in March 2018. The study was approved by the Institutional Review Board of our university. 

\subsubsection{Participants}

We recruited a total of 1,052 participants in the YouTube experiment and 753 participants in the Pinterest experiment, resulting in $\sim$30 participants in each condition for each video and pin. Of the 1,052 participants in the YouTube experiment, 14 ($\sim$1.3\%) skipped watching the video---as revealed by our measurements---or were not paying attention---as revealed by their open-ended responses---and were subsequently filtered out. 

The mean age of participants in the YouTube experiment was 37.78, with a standard deviation of 12.11. Over half (52.60\%) of all these participants were male, and nearly two-thirds (64.07\%) had either a college or bachelor's degree. Close to nine-tenths (91.62\%) of all participants had a YouTube account. Similarly, the mean age of participants in the Pinterest experiment was 38.29, with a standard deviation of 12.49. A little over two-fifths (41.57\%) of all participants were male, and nearly two-thirds (63.21\%) had either a college or bachelor's degree. Close to fourth-fifths (82.50\%) of all participants had a Pinterest account.

\subsubsection{Data Analysis}

We built ordinal logistic regression mixed-models to analyze participant responses. Ordinal regression is used when the dependent variable is ordinal in nature (e.g., a Likert scale response). We used mixed-models to account for the \emph{video/pin} independent variable, which we considered as a random factor. We built separate models for both YouTube and Pinterest using the ordinal package in R\footnote{https://cran.r-project.org/web/packages/ordinal/index.html}. For each model, we also verified that the proportional odds assumption was not violated.

To analyze the open-ended response in which we asked participants to explain what the disclosure statement meant to them (\emph{explain}), we adopted a qualitative data analysis approach and performed deductive coding~\cite{seidman2013interviewing}. Two researchers created and agreed upon a codebook after an initial exploration of all the responses. Using this codebook, both researchers independently coded the responses, blind to the condition of the responses to avoid biasing the assigned codes. We then calculated inter-rater reliability using Krippendorff's alpha ($\alpha$), and achieved an $\alpha = 0.82$ and $\alpha = 0.84$ in the YouTube and Pinterest experiment responses, indicating high agreement. We then resolved the disagreements, and arrived at the final codes for each response. We make our codebook available in the Appendix (Table~\ref{tab:codebook}).

\subsubsection{Limitations}

Our user study has limitations. First, we only tested one disclosure statement within each disclosure type on both YouTube and Pinterest. We picked statements that were fairly generic enough to generalize across all the videos and pins. Second, our studies were limited in generalizability to the broader MTurk population. While the demographics of our participants, including age, gender, and education, were broadly comparable to the Internet users population~\cite{pew}, MTurk users may have differed in terms of Internet literacy, proficiency, and experience. Third, we randomly sampled the five YouTube videos around the median length of affiliate videos, and as a result our results may not generalize to videos of relatively longer and shorter duration.

\subsection{Findings}
In the following sections, we present our findings, highlighting the efficacy of the various disclosure types on both platforms. The output of the various ordinal logistic regression models for the YouTube and Pinterest experiments is summarized in Table~\ref{tab:one} and Table~\ref{tab:two} respectively. For each dependent variable across both experiments, the regression models compared the treatment conditions against the control.

\begin{figure}[t]
    \centering
    \begin{subfigure}[b]{0.47\textwidth}
        \includegraphics[width=\textwidth]{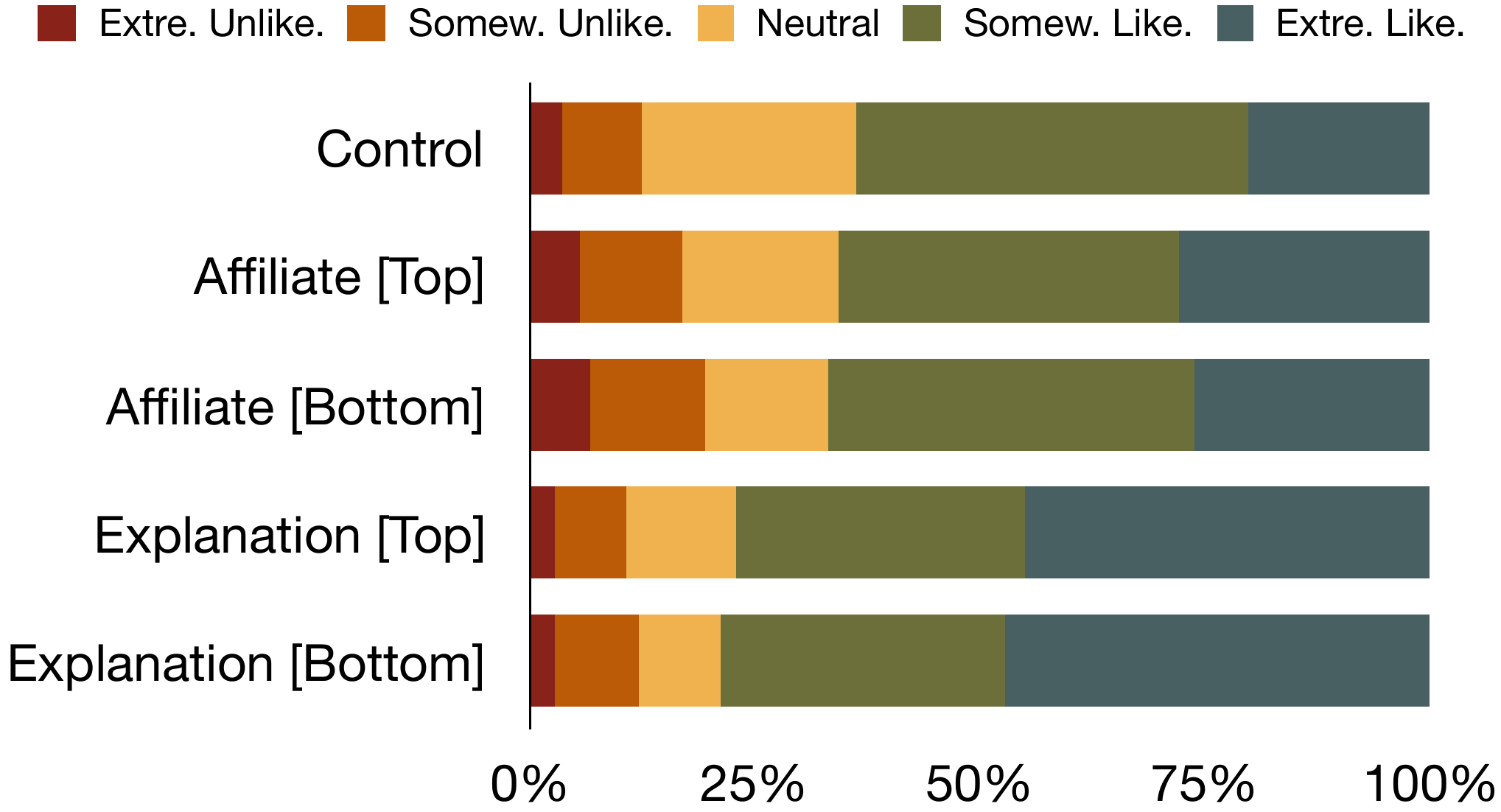}
        \caption{Content Creator Relationship}
        \label{c1}
    \end{subfigure}
    \begin{subfigure}[b]{0.47\textwidth}
        \includegraphics[width=\textwidth]{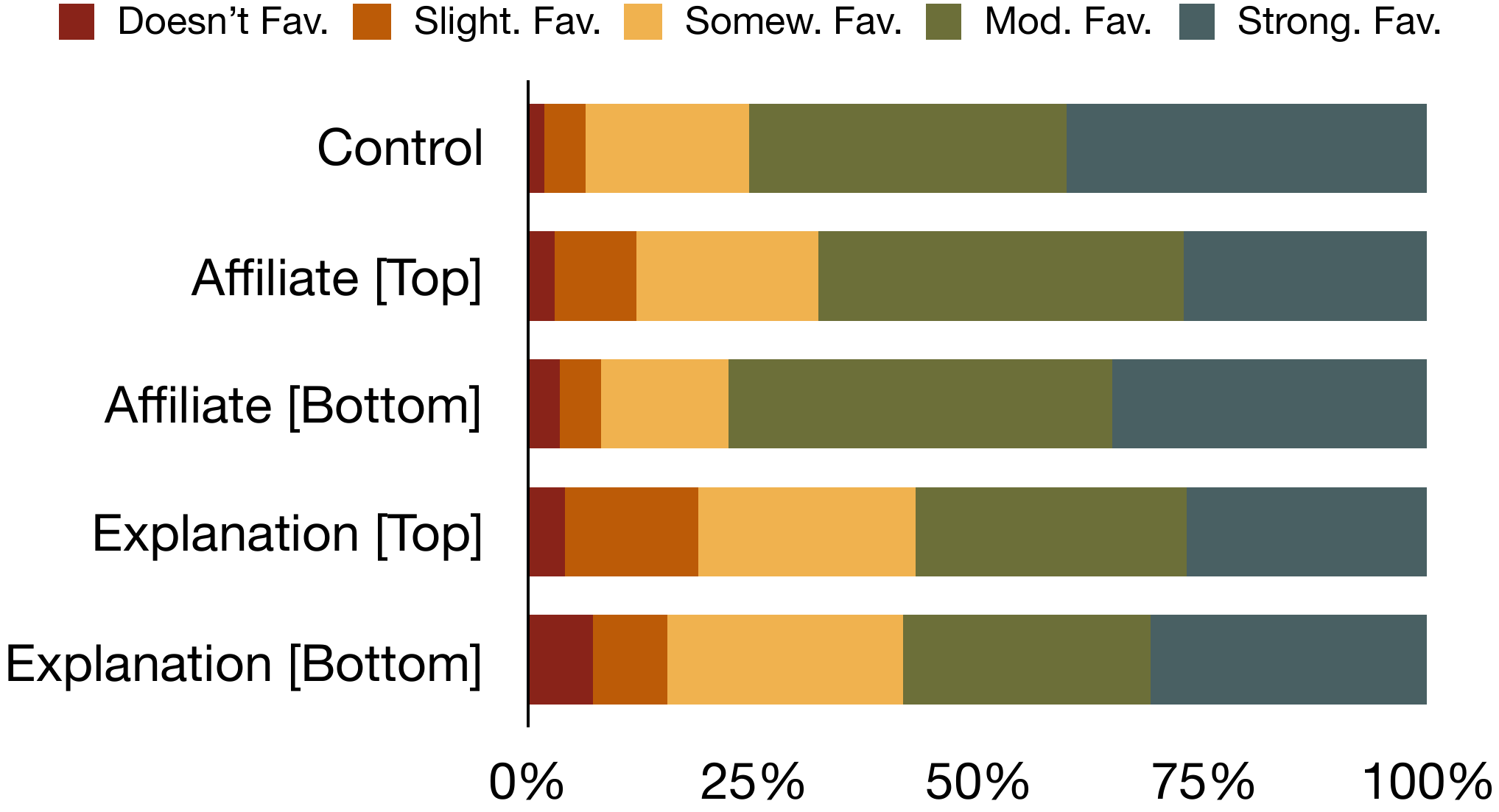}
        \caption{Content Creator Favors}
        \label{c2}
    \end{subfigure}
    \caption{Distribution of the \emph{Creator Relationship} and \emph{Creator Favors} Variables from the Pinterest Experiment.}
    \label{fig:chart}
\end{figure}

\subsubsection{Users Notice Explanation Disclosures, But Only On Pinterest}

We asked participants how likely it was that there was a relationship between the content creator and the merchant selling the product. As seen in Table~\ref{tab:one} we found no evidence that suggested any of the disclosure types in the YouTube experiment had a statistically significant effect on participants' awareness of this relationship when compared to the control condition. This suggests that the participants either failed to notice the disclosures or interpreted them incorrectly, regardless of their type and position.

However, in the Pinterest experiment, and as summarized in Table~\ref{tab:two}, we found that the presence of a \emph{Explanation} disclosure either at the top ($O.R. = 2.37$, $p < 0.0001$) or bottom ($O.R. = 2.82$, $p < 0.0001$) of the description multiplicatively increased the odds of participants' awareness of the relationship between the content creator and the merchant selling the product by 2.37 and 2.82 times respectively when compared to the control condition. Figure~\ref{c1} shows the distribution of the \emph{content creator relationship} dependent variable aggregated across the pins. While about 60\% of participants in the control condition stated that it was either somewhat or extremely likely that there was a relationship between the content creator and the merchant, nearly 75\% did so in the \emph{Explanation} disclosure (Top \& Bottom) conditions. This suggests that participants noticed and correctly interpreted the \emph{Explanation} disclosures regardless of their position in the pins' description. 

Across both platforms, we found no evidence to support that \emph{Affiliate Link} disclosures were effective. As we show in a later result, this is likely not just because users may have not noticed these disclosure, but also due to unclear nature of the disclosure wording itself.

\subsubsection{Similar Product Impressions Across All Conditions}

We asked participants about their impression of the videos and pins (\emph{video, pin impression}) before they were exposed to any treatment. As shown in Table~\ref{tab:one} and Table~\ref{tab:two}, we found no evidence that suggested any of the disclosure types had a statistically significant effect on participants' impressions of the videos and pins when compared to the control condition in both experiments. We expected this result because the participants had not yet been exposed to the treatments when they answered this question. Therefore, any impressions of the videos and pins were likely averaged out due to randomization.

After participants read through the description of the video and pin---which varied depending on the condition of the experiments---we asked them about their impression of the product (\emph{product impression}) that was linked by the content creator using an affiliate URL. As shown in Table~\ref{tab:one} and Table~\ref{tab:two}, we found no evidence that suggested any of the disclosure types had a statistically significant effect on participants' impressions of the product when compared to the control condition in both the YouTube or Pinterest experiments. This suggests that participants' impression of the products did not vary due to the presence of any of the kinds of disclosures.

\begin{table}%
\captionsetup{justification=centering}
\caption{Ordinal Logistic Regression Models for the YouTube Experiment. Each Column Represents an Ordinal Regression Model. The Name of the Column Indicates the Dependent Variable. Values Within Brackets Represent the 95\% Confidence Interval.\\ 
\colorbox[rgb]{0.87,0.92,0.97}{$\ast$} $p < 0.05\enspace$ \colorbox[rgb]{0.78,0.86,0.94}{$\ast\ast$} $p < 0.01\enspace$ \colorbox[rgb]{0.62,0.79,0.88}{$\ast\ast\ast$} $p < 0.001\enspace$ \colorbox[rgb]{0.42,0.68,0.84}{$\ast\ast\ast\ast$} $p < 0.0001\enspace$}
\label{tab:one}
\begin{minipage}{\columnwidth}
\begin{center}

\begin{tabular}{@{}lcccc@{}}
\toprule
\multicolumn{1}{c}{\textbf{\begin{tabular}[c]{@{}c@{}}Experimental\\ Condition(s)\end{tabular}}} & \textbf{\begin{tabular}[c]{@{}c@{}}Video\\ Impression\end{tabular}} & \textbf{\begin{tabular}[c]{@{}c@{}}Product\\ Impression\end{tabular}} & \textbf{\begin{tabular}[c]{@{}c@{}}Creator\\ Favors\end{tabular}} & \textbf{\begin{tabular}[c]{@{}c@{}}Creator\\ Relationship\end{tabular}} \\
\multicolumn{1}{c}{{[}Control: No Disclosure{]}}                                                 & \multicolumn{4}{c}{Odds Estimate {[}95\% C.I.{]}}                                                                                                                                                                                                                                      \\ \midrule
\begin{tabular}[c]{@{}l@{}}Affiliate Link {[}Top{]}\end{tabular}                               & \begin{tabular}[c]{@{}c@{}}0.90\\ {[}0.60, 1.36{]}\end{tabular}     & \begin{tabular}[c]{@{}c@{}}0.85\\ {[}0.56, 1.29{]}\end{tabular}    & \begin{tabular}[c]{@{}c@{}}0.78\\ {[}0.52, 1.17{]}\end{tabular}   & \begin{tabular}[c]{@{}c@{}}0.79\\ {[}0.53, 1.17{]}\end{tabular}         \\\rule{0pt}{4ex}
\begin{tabular}[c]{@{}l@{}}Affiliate Link {[}Bottom{]}\end{tabular}                            & \begin{tabular}[c]{@{}c@{}}0.81\\ {[}0.53, 1.23{]}\end{tabular}     & \begin{tabular}[c]{@{}c@{}}0.91\\ {[}0.59, 1.40{]}\end{tabular}    & \begin{tabular}[c]{@{}c@{}}1.01\\ {[}0.68, 1.55{]}\end{tabular}   & \begin{tabular}[c]{@{}c@{}}1.00\\ {[}0.66, 1.52{]}\end{tabular}         \\\rule{0pt}{4ex}
\begin{tabular}[c]{@{}l@{}}Explanation {[}Top{]}\end{tabular}                         & \begin{tabular}[c]{@{}c@{}}0.74\\ {[}0.48, 1.13{]}\end{tabular}     & \begin{tabular}[c]{@{}c@{}}0.89\\ {[}0.58, 1.38{]}\end{tabular}    & \begin{tabular}[c]{@{}c@{}}0.80\\ {[}0.52, 1.23{]}\end{tabular}   & \begin{tabular}[c]{@{}c@{}}1.08\\ {[}0.71, 1.66{]}\end{tabular}         \\\rule{0pt}{4ex}
\begin{tabular}[c]{@{}l@{}}Explanation {[}Bottom{]}\end{tabular}                      & \begin{tabular}[c]{@{}c@{}}0.86\\ {[}0.56, 1.32{]}\end{tabular}     & \begin{tabular}[c]{@{}c@{}}0.91\\ {[}0.63, 1.52{]}\end{tabular}    & \begin{tabular}[c]{@{}c@{}}0.84\\ {[}0.54, 1.30{]}\end{tabular}   & \begin{tabular}[c]{@{}c@{}}1.30\\ {[}0.85, 1.97{]}\end{tabular}         \\\rule{0pt}{4ex}
\begin{tabular}[c]{@{}l@{}}Channel Support {[}Top{]}\end{tabular}                              & \begin{tabular}[c]{@{}c@{}}0.86\\ {[}0.55, 1.33{]}\end{tabular}     & \begin{tabular}[c]{@{}c@{}}0.73\\ {[}0.47, 1.14{]}\end{tabular}    & \begin{tabular}[c]{@{}c@{}}0.89\\ {[}0.57, 1.39{]}\end{tabular}   & \begin{tabular}[c]{@{}c@{}}0.68\\ {[}0.44, 1.05{]}\end{tabular}         \\\rule{0pt}{4ex}
\begin{tabular}[c]{@{}l@{}}Channel Support {[}Bottom{]}\end{tabular}                           & \begin{tabular}[c]{@{}c@{}}0.91\\ {[}0.60, 1.39{]}\end{tabular}     & \begin{tabular}[c]{@{}c@{}}0.95\\ {[}0.62, 1.45{]}\end{tabular}    & \begin{tabular}[c]{@{}c@{}}0.93\\ {[}0.61, 1.42{]}\end{tabular}   & \begin{tabular}[c]{@{}c@{}}0.74\\ {[}0.49, 1.12{]}\end{tabular}         \\\rule{0pt}{4ex}
\begin{tabular}[c]{@{}l@{}}Random Effect {[}Std. Dev.{]}\end{tabular}                          & 0.67                                                                & 0.66                                                               & 0.68                                                              & 0.42                                                                    \\ \bottomrule
\end{tabular}

\end{center}
\end{minipage}
\end{table}%

\subsubsection{Explanation Disclosures Decrease Participants' Perceptions of Content Creators' Endorsement On Pinterest}

After they had read through the videos' and pins' descriptions, we asked participants how much they thought the content creator favored the product that was linked using an affiliate URL (\emph{content creator favors}). Only 3.13\% and 3.71\% of participants---across all the conditions---in the YouTube and Pinterest experiments respectively thought the content creator did not favor the product at all. Specifically in the control condition---having no disclosure---this number was comparable: only 2.76\% and 1.75\% of all participants in the YouTube and Pinterest experiments respectively thought the content creator did not favor the product at all. This suggests that participants do think of the products featured through affiliate URLs as endorsements by the creators.

As seen in Table~\ref{tab:one}, we found no evidence that suggested participants' perceptions of the creators' endorsement of the affiliate marketing product varied across the disclosure types in the YouTube experiment compared to the control condition. However, in the Pinterest experiment, and as summarized in Table~\ref{tab:two}, we found that the presence of an \emph{Affiliate Link} disclosure at the top of the description ($O.R. = 0.62$, $p < 0.05$), or an \emph{Explanation} disclosure both at the top ($O.R. = 0.46$, $p < 0.001$) or the bottom ($O.R. = 0.53$, $p < 0.01$) of the description, all multiplicatively decreased the odds of participants' perceptions of content creators' endorsement of the product by nearly half when compared to the control condition. Figure~\ref{c2} shows the distribution of the \emph{content creator favors} dependent variable aggregated across the pins. While about 75\% of participants in the control condition stated that the content creator either moderately or strongly favored the product, only 50\% of participants did so in the \emph{Explanation} (Top \& Bottom) conditions. This result from the Pinterest experiment suggests that participants noticed and correctly interpreted the \emph{Explanation} disclosure, and as a result perceived the content creators' disposition to be more neutral relative to the control condition. As before, we found no evidence to support that \emph{Affiliate Link} disclosures were effective.

\begin{table}%
\captionsetup{justification=centering}
\caption{Ordinal Logistic Regression Models for the Pinterest Experiment. Each Column Represents an Ordinal Regression Model. The Name of the Column Indicates the Dependent Variable. Values Within Brackets Represent the 95\% Confidence Interval.\\
\colorbox[rgb]{0.87,0.92,0.97}{$\ast$} $p < 0.05\enspace$ \colorbox[rgb]{0.78,0.86,0.94}{$\ast\ast$} $p < 0.01\enspace$ \colorbox[rgb]{0.62,0.79,0.88}{$\ast\ast\ast$} $p < 0.001\enspace$ \colorbox[rgb]{0.42,0.68,0.84}{$\ast\ast\ast\ast$} $p < 0.0001\enspace$}
\label{tab:two}
\begin{minipage}{\columnwidth}
\begin{center}
\begin{tabular}{@{}lcccc@{}}
\toprule
\multicolumn{1}{c}{\textbf{\begin{tabular}[c]{@{}c@{}}Experimental\\ Condition(s)\end{tabular}}} & \textbf{\begin{tabular}[c]{@{}c@{}}Pin\\ Impression\end{tabular}} & \textbf{\begin{tabular}[c]{@{}c@{}}Product\\ Impression\end{tabular}} & \textbf{\begin{tabular}[c]{@{}c@{}}Creator\\ Favors\end{tabular}}                       & \textbf{\begin{tabular}[c]{@{}c@{}}Creator\\ Relationship\end{tabular}}                 \\
\multicolumn{1}{c}{{[}Control: No Disclosure{]}}                                                 & \multicolumn{4}{c}{Odds Estimate {[}95\% C.I.{]}}                                                                                                                                                                                                                                                                          \\ \midrule
\begin{tabular}[c]{@{}l@{}}Affiliate Link {[}Top{]}\end{tabular}                               & \begin{tabular}[c]{@{}c@{}}0.98\\ {[}0.65, 1.47{]}\end{tabular}   & \begin{tabular}[c]{@{}c@{}}1.49\\ {[}0.99, 2.24{]}\end{tabular}    & \cellcolor[HTML]{DEEBF7}\begin{tabular}[c]{@{}c@{}}$0.62^{\ast}$\\ {[}0.42, 0.93{]}\end{tabular} & \begin{tabular}[c]{@{}c@{}}1.21\\ {[}0.81, 1.81{]}\end{tabular}                         \\ \rule{0pt}{4ex}
\begin{tabular}[c]{@{}l@{}}Affiliate Link {[}Bottom{]}\end{tabular}                            & \begin{tabular}[c]{@{}c@{}}0.92\\ {[}0.61, 1.38{]}\end{tabular}   & \begin{tabular}[c]{@{}c@{}}1.44\\ {[}0.96, 2.16{]}\end{tabular}    & \begin{tabular}[c]{@{}c@{}}0.90\\ {[}0.61, 1.35{]}\end{tabular}                         & \begin{tabular}[c]{@{}c@{}}1.13\\ {[}0.76, 1.68{]}\end{tabular}                         \\ \rule{0pt}{4ex}
\begin{tabular}[c]{@{}l@{}}Explanation {[}Top{]}\end{tabular}                         & \begin{tabular}[c]{@{}c@{}}1.01\\ {[}0.67, 1.52{]}\end{tabular}   & \begin{tabular}[c]{@{}c@{}}0.93\\ {[}0.62, 1.40{]}\end{tabular}    & \cellcolor[HTML]{9ECAE1}\begin{tabular}[c]{@{}c@{}}$0.46^{\ast\ast\ast}$\\ {[}0.30, 0.69{]}\end{tabular} & \cellcolor[HTML]{6BAED6}\begin{tabular}[c]{@{}c@{}}$2.37^{\ast\ast\ast\ast}$\\ {[}1.58, 3.56{]}\end{tabular} \\ \rule{0pt}{4ex}
\begin{tabular}[c]{@{}l@{}}Explanation {[}Bottom{]}\end{tabular}                      & \begin{tabular}[c]{@{}c@{}}1.29\\ {[}0.85, 1.87{]}\end{tabular}   & \begin{tabular}[c]{@{}c@{}}1.24\\ {[}0.82, 1.87{]}\end{tabular}    & \cellcolor[HTML]{C6DBEF}\begin{tabular}[c]{@{}c@{}}$0.53^{\ast\ast}$\\ {[}0.35, 0.80{]}\end{tabular} & \cellcolor[HTML]{6BAED6}\begin{tabular}[c]{@{}c@{}}$2.82^{\ast\ast\ast\ast}$\\ {[}1.86, 4.29{]}\end{tabular} \\ \rule{0pt}{4ex}
\begin{tabular}[c]{@{}l@{}}Random Effect {[}Std. Dev.{]}\end{tabular}                          & 0.49                                                              & 0.25                                                               & 0.12                                                                                    & 0.44                                                                                    \\ \bottomrule
\end{tabular}

\end{center}
\end{minipage}
\end{table}%

\begin{table}
\centering
\caption{Participants' Interpretation of the Disclosure Statement Broken Down by Disclosure Type.}
\label{tab:expl}
\begin{tabular}{lcccc}
\toprule
\multicolumn{1}{c}{\multirow{2}{*}{\textbf{\begin{tabular}[c]{@{}c@{}}Disclosure \\ Type\end{tabular}}}} & \multicolumn{2}{c}{\textbf{YouTube}}           & \multicolumn{2}{c}{\textbf{Pinterest}}         \\ \cline{2-5} 
\multicolumn{1}{c}{}                                                                                     & \textbf{Incorrect}      & \textbf{Correct}     & \textbf{Incorrect}      & \textbf{Correct}     \\ \midrule
Affiliate Link                                                                                           & 115                     & 217                  & 145                     & 146                  \\
Explanation                                                                                              & 14                      & 270                  & 15                      & 276                  \\
Channel Support                                                                                          & 40                      & 236                  & NA                      & NA            \\ \bottomrule    
\end{tabular}
\end{table}

\subsubsection{Participants Accurately Interpret Explanation Disclosures}

We asked participants to explain what the disclosure statement meant to them in their own words by means of an open-ended question. We classified the resulting codes from the qualitative data analysis into two groups: \emph{Correct} and \emph{Incorrect}. The first group consisted of all those codes that suggested a relationship---not just financial---between the content creator and the merchant, or that the content was an advertisement, was sponsored or was a promotion. The second group consisted of all other codes; these represented either incorrect interpretations of the disclosure statements or null responses. Our goal with this division of codes was not motivated by whether participants understood the specifics of affiliate marketing, but more generally, that they understood that the content creator benefited in some way from a partnership with the merchant. Table~\ref{tab:codebook} in the Appendix indicates the codes we marked \emph{Correct} and \emph{Incorrect}.

Table~\ref{tab:expl} summarizes our results. In the YouTube experiment, nearly 95\%, 85\%, and only 65\% of participants in the \emph{Explanation}, \emph{Channel Support}, and \emph{Affiliate Link} conditions respectively correctly interpreted the disclosure statement. A Chi-sq test of independence was significant (${\chi}^2 = 93.14$, $p < 0.0001$), indicating indicating that the disclosure types and their interpretation---as defined by our groups---were not independent of each other. Post-hoc comparison tests revealed that there was a statically significant difference between the \emph{Affiliate Link} and \emph{Explanation} conditions (${\chi}^2 = 79.82$, $p < 0.0001$), the \emph{Affiliate Link} and \emph{Channel Support} conditions (${\chi}^2 = 31.15$, $p < 0.0001$), and the \emph{Explanation} and \emph{Channel Support} conditions (${\chi}^2 = 13.62$, $p < 0.001$). Similarly, in the Pinterest experiment, nearly 95\% and only 50\%  of participants in the \emph{Explanation} and \emph{Affiliate Link} conditions respectively correctly interpreted the disclosure statement. A Chi-square test of independence was---as before---significant (${\chi}^2 = 183.85$, $p < 0.0001$).

Overall, when presented with the disclosure statement, the majority of participants correctly interpreted the \emph{Explanation} and \emph{Channel Support} disclosures, understanding that posting the affiliate marketing content benefited the content creator in some way. In contrast, fewer participants found \emph{Affiliate Link} disclosures interpretable. 

In summary, we found that \emph{Explanation} disclosures are more effective than \emph{Affiliate Link} disclosures when users can find them (e.g., on Pinterest); users likely fail to either notice or correctly interpret \emph{Affiliate Link} disclosures. Further, we think the \emph{Explanation} disclosures likely had an effect in the Pinterest experiment but not the YouTube experiment because of the norms and design of social media platforms. In the five affiliate pins we used in the experiment---and this is true for affiliate Pinterest pins more generally (\emph{Median = 1 line})---the descriptions were short and no more than one or two lines. On the other hand, because YouTube descriptions in the five videos were longer---with this being true for affiliate YouTube videos more generally (\emph{Median = 14 lines})---this may have limited the ability of users to identify the disclosures.

\section{Discussion}
In this section, we discuss the implications of our findings and suggest directions for future work.

\subsection{Detecting Affiliate Marketing Content and Disclosures}
Through our work, we demonstrated a new method for discovering affiliate marketing companies' URL patterns. Using this method, we were able to detect affiliate marketing content on two social media platforms and generate a publicly available list for researchers to contribute to and use. Similarly, our method for discovering disclosures within content creators' textual descriptions can aid in building automated disclosure detection tools; we think this can be of broad use for regulators, platforms, and researchers to ultimately help users better identify endorsement-based advertisements. Lastly, both methods are extensible and can be used to replicate our study on other social media platforms.

\subsection{Effectiveness of the FTC's Endorsement Guidelines}

\subsubsection{Low Prevalence of Disclosures}

Our results showed that only a small percentage of affiliate marketing content from YouTube and Pinterest contained any disclosures at all. This suggests that the FTC's endorsement guidelines have only had a limited effect in ensuring that content creators disclose their affiliate marketing relationship to consumers. Even within the small set of disclosures that are made, the majority are of the kind the FTC specifically discourages using: the short \emph{Affiliate Link} disclosures.
 
This presents an opportunity for further investigation and study: why do content creators fail to follow the guidelines? We think there might be at least two contributing reasons. First, it is not clear whether content creators are even aware of the guidelines to begin with. Future work could---by means of surveys and interviews---examine the reasons behind why content creators on social media choose to or fail to disclose their affiliate marketing content. Second, the FTC has only recently begun enforcing its endorsement guidelines, setting precedents for violations~\cite{ftc1, ftc2, ftc3, ftc4}. We expect that endorsement disclosures---including those related to affiliate marketing---will appear more frequently as awareness around the guidelines grows.

\subsubsection{Explanatory Disclosures Help But Only When Users Can Find Them}

Our study validates the FTC's endorsement guidelines that explanatory disclosures are indeed effective. Users are able to better understand the underlying relationship between the content creator and the merchants through the \emph{Explanation} disclosure, but our results also show platform-specific effects: users may not be able to notice and interpret disclosures that are buried in long texts, regardless of their clarity, such as on YouTube. Therefore, we recommend that content creators ensure that they create disclosures in a manner that facilitates discovery (e.g., by surrounding the disclosure by asterisks and underscores). 

\subsection{Shifting the Burden to Affiliate Marketing Companies}

Content creators could be held accountable to better disclosure practices by the affiliate marketing companies they associate with. We examined the affiliate marketing terms and conditions, where publicly available, of eight of the most prevalent affiliate marketing companies from our dataset: Amazon, AliExpress, Commission Junction, Rakuten Marketing, Impact Radius, RewardStyle, ShopStyle and ShareASale. We could not find any publicly available terms and conditions from Impact Radius or RewardStyle. Only Amazon\footnote{https://affiliate-program.amazon.com/help/operating/agreement} and ShopStyle\footnote{https://www.shopstylecollective.com/terms} explicitly referenced the FTC's guidelines in their affiliate terms. While we did not find any references in the Rakuten Marketing, ShareASale, Commission Junction affiliate terms, we noted all three companies blogging about the guidelines on their company blogs\footnote{http://blog.shareasale.com/2017/09/08/ftc-updates-and-faq-s/}\textsuperscript{,}\footnote{https://blog.marketing.rakuten.com/topic/ftc-disclosure-guidelines}\textsuperscript{,}\footnote{http://junction.cj.com/article/disclosures-what-you-need-know}.

However, its unclear how many other affiliate marketing companies follow this practice, and whether they explicitly point the content creators that register with them to the FTC's guidelines. In addition to focusing on the content creators, we suggest that the FTC could work with these companies to revise their disclosure requirements and standards.

\subsection{Design Suggestions for Effective Disclosures}

Our findings also unlock several design suggestions that various stakeholders in the social media advertising ecosystem---including social media platforms and web browsers---can incorporate to improve disclosure practices, and help users identify these disclosures.

\subsubsection{Affordances Through Social Media Platforms}

Along with content creators, social media platforms play a critical role in shaping the disclosures that content creators make. The disclosures, generally speaking, are limited by the character space available to them. For instance, the description length can be as long as 5,000 characters on YouTube, but on Pinterest it is capped to 500 characters; similarly, tweets can only be as long as 280 characters on Twitter. Because some of these interfaces may be more more conducive for content creators to add disclosures than others, social media platforms can help design their interfaces to make it easier for content creators to disclose without crowding their other promotion text.

In fact, Instagram recently added an option for sponsored content to be disclosed by using the \emph{Paid partnership} tool, which enables disclosures outside of the traditional image description \cite{instagram}. Similarly, YouTube recently added the ability for create a \emph{Contains Product Placement} overlay to their videos \cite{youtube}. Such disclosure tools are a step in the right direction, however it is unlikely that any one blanket disclosure will cover all marketing practices. As the FTC's endorsement guidelines evolve, social media platforms can also help design tools that enable content creators to retroactively edit disclosures from their past posts. Future research could focus on examining in depth the kinds of affordances that can be built into social media platforms to enable content creators to disclose their advertisements clearly and conspicuously.

\subsubsection{Protection Through Web Browsers}

Web browsers could also help inform users that the content they are watching has affiliate URLs by means of in-built support or through add-ons and extension. Such tools could function like current Ad-blockers, but rather than blocking advertisements, they could either highlight when a piece of content should contain disclosures, or highlight the actual disclosures when they are present. These tools could leverage \emph{Explanation} disclosures' high interpretability rate in presenting the disclosure information to users.

Web browsers can achieve this in practice using machine learning and natural language processing based approaches in which algorithms can be trained on large datasets of labelled social media content. Our findings provide a starting point, showing that the three categories of affiliate disclosures often use wording that exhibit a common pattern. In future work, we aim to build such tools, and conduct user studies to evaluate how well they work in practice.



\section{Conclusion}
We examined advertising disclosures accompanying affiliate marketing content on two social media platforms: YouTube and Pinterest. Our study reveals that only about 10\% of affiliate marketing content on both platforms contains \emph{any} disclosures at all. Further, the explanatory disclosures which the FTC recommends using are indeed the ones we found most effective; however, these were also the least prevalent of all disclosure types. We offer practical recommendations from both a design and policy perspective to help enable better disclosures for users. Doing so can ultimately help increase transparency surrounding social media endorsements.


\begin{acks}
We thank our participants for taking part in our study. We also thank Nathan Malkin and the anonymous reviewers for their feedback on our paper. This research was funded by NSF grant CNS-1664786.
\end{acks}

\appendix

\begin{table*}[ht]
\tiny
  \centering
  \caption{List of Affiliate Marketing Companies Discovered by Our Analysis. Count Indicates the Number of times the URL Pattern Appeared When We Resolved the Retrieved URLs.}
\label{tab:affiliate}
\begin{tabular}{l l l r r}
\toprule
\textbf{Company Name} & \textbf{Domain} & \textbf{URL Pattern} & \textbf{YouTube Count} & \textbf{Pinterest Count}\\

\midrule
\multirow{2}{*}{admitad}  & \multirow{2}{*}{admitad} & https://ad.admitad.com/g/\ldots & \multirow{2}{*}{245}  & \multirow{2}{*}{1} \\
&  & https://ad.admitad.com/goto/\ldots & &  \\[0.12cm]

affiliaXe  & affiliaxe & http://performance.affiliaxe.com/\ldots\&aff\_id=\ldots & 151 & 0 \\[0.12cm]

AliExpress & aliexpress & https://s.aliexpress.com/\ldots\&af=\ldots & 2167 & 785 \\[0.12cm]

Amazon & amazon & http://www.amazon.(com,de,fr,in,it)/\ldots\&tag=\ldots & 7308 & 7368 \\[0.12cm]

Apple & apple & https://itunes.apple.com/\ldots\&at=\ldots & 669 & 61 \\[0.12cm]

Audiobooks & audiobooks & https://affiliates.audiobooks.com/\ldots\&a\_aid=\ldots\&a\_bid=\ldots & 129 & 0 \\[0.12cm]

AvantLink & avantlink & http://www.avantlink.com/\ldots\&pw=\ldots & 34 & 12 \\[0.12cm]

Avangate & avangate & https://secure.avangate.com/\ldots\&AFFILIATE=\ldots & 12 & 0 \\[0.12cm]

\multirow{3}{*}{Awin} & awin1 & http://www.awin1.com/\ldots\&awinaffid=\ldots & \multirow{3}{*}{129} & \multirow{3}{*}{211} \\
& zanox & http://ad.zanox.com/ppc/?\ldots &  & \\
& zenaps & http://www.zenaps.com/rclick.php?\ldots &  & \\[0.12cm]

Banggood & banggood & http://www.banggood.com/\ldots\&p=\ldots & 88 & 13\\[0.12cm]

Book Depository & bookdepository & https://www.bookdepository.com/\ldots\&a\_aid=\ldots & 103 & 0\\[0.12cm]

Booking.com & booking & https://www.booking.com/\ldots\&aid=\ldots & 257 & 7\\[0.12cm]

Clickbank & clickbank & http://\ldots .hop.clickbank.net/\ldots & 678 & 262 \\[0.12cm]

\multirow{8}{*}{CJ Affiliate}  & anrdoezrs & http://www.anrdoezrs.net/click-[0-9]+-[0-9]+\ldots & \multirow{8}{*}{341} & \multirow{8}{*}{2413} \\
 & dotomi & http://cj.dotomi.com/\ldots &  &  \\
 & dpbolvw & http://www.dpbolvw.net/click-[0-9]+-[0-9]+\ldots &  &  \\
 & emjcd & http://www.emjcd.com/\ldots &  &  \\
 & jdoqocy & http://www.jdoqocy.com/click-[0-9]+-[0-9]+\ldots &  &  \\
 & kqzyfj & http://www.kqzyfj.com/click-[0-9]+-[0-9]+\ldots &  &  \\
 & qksrv & http://qksrv.net/\ldots &  &  \\
 & tkqlhce & http://www.tkqlhce.com/click-[0-9]+-[0-9]+\ldots &  &  \\[0.12cm]

Ebay & ebay & http://rover.ebay.com\ldots\&campid=\ldots & 99 & 1963\\[0.12cm]

\multirow{6}{*}{Envato}  & audiojungle & https://audiojungle.net/\ldots\&ref=\ldots & 108 & 0 \\
& codecanyon & https://codecanyon.net/\ldots\&ref=\ldots & 14 & 76 \\
& envato & https://marketplace.envato.com/\ldots\&ref=\ldots & 175 & 262 \\
& graphicriver & https://graphicriver.net/\ldots\&ref=\ldots & 15 & 1465 \\
& themeforest & https://themeforest.net/\ldots\&ref=\ldots & 19 & 200 \\
& videohive & https://videohive.net/\ldots\&ref=\ldots & 578 & 33 \\[0.12cm]

e-Commerce & \multirow{2}{*}{buyeasy} & http://buyeasy.by/cashback/\ldots & \multirow{2}{*}{741} & \multirow{2}{*}{7} \\
Partners Network &  & http://buyeasy.by/redirect/\ldots & & \\[0.12cm]

Flipkart & flipkart & https://www.flipkart.com/\ldots\&affid=\ldots & 81 & 20 \\[0.12cm]

GT Omega Racing & gtomegaracing & http://www.gtomegaracing.com/\ldots\&tracking=\ldots & 56 & 0 \\[0.12cm]

Hotellook & hotellook & https://search.hotellook.com/\ldots\&marker=\ldots & 165 & 5 \\[0.12cm]

Hotmart & hotmart & https://www.hotmart.net.br/\ldots\&a=\ldots & 211 & 8 \\[0.12cm]

\multirow{2}{*}{Impact Radius} & 7eer & http://\ldots.7eer.net/c/[0-9]+/[0-9]+/[0-9]+\ldots & \multirow{2}{*}{180} & \multirow{2}{*}{529} \\
& evyy & http://\ldots.evyy.net/c/[0-9]+/[0-9]+/[0-9]+\ldots & & \\[0.12cm]

KontrolFreek & kontrolfreek & https://www.kontrolfreek.com/\ldots\&a\_aid=\ldots & 117 & 0 \\[0.12cm]

Makeup Geek & makeupgeek & http://www.makeupgeek.com/\ldots\&acc=\ldots & 57 & 0 \\[0.12cm]

\multirow{6}{*}{Pepperjam Network} & gopjn & http://www.gopjn.com/t/[0-9]-[0-9]+-[0-9]+-[0-9]+\ldots & \multirow{6}{*}{2} & \multirow{6}{*}{79} \\
& pjatr & http://www.pjatr.com/t/[0-9]-[0-9]+-[0-9]+-[0-9]+\ldots &  &  \\
 & pjtra & http://www.pjtra.com/t/[0-9]-[0-9]+-[0-9]+-[0-9]+\ldots &  &  \\
 & pntra & http://www.pntra.com/t/[0-9]-[0-9]+-[0-9]+-[0-9]+\ldots &  &  \\
 & pntrac & http://www.pntrac.com/t/[0-9]-[0-9]+-[0-9]+-[0-9]+\ldots &  &  \\
 & pntrs & http://www.pntrs.com/t/[0-9]-[0-9]+-[0-9]+-[0-9]+\ldots &  &  \\[0.12cm]

Rakuten Marketing & linksynergy & http://click.linksynergy.com/\ldots\&id=\ldots & 189 & 1877 \\[0.12cm]

Skimlinks & redirectingat & http://go.redirectingat.com/\ldots\&id=\ldots & 43 & 155 \\[0.12cm]

Smartex & olymptrade & https://olymptrade.com/\ldots\&affiliate\_id=\ldots & 65 & 0 \\[0.12cm]

RewardStyle & rstyle & http://rstyle.me/\ldots & 402 & 2711 \\[0.12cm]

ShopStyle & shopstyle & http://shopstyle.it/\ldots & 111 & 9239 \\[0.12cm]

\multirow{3}{*}{ShareASale} & \multirow{3}{*}{shareasale} & http://www.shareasale.com/r.cfm\ldots & \multirow{3}{*}{199} & \multirow{3}{*}{616} \\
 &  & http://www.shareasale.com/m-pr.cfm\ldots &  &  \\
 &  & http://www.shareasale.com/u.cfm\ldots &  &  \\[0.12cm]

Studybay & apessay & https://apessay.com/\ldots\&rid=\ldots & 141 & 0 \\[0.12cm]

Zaful & zaful & http://zaful.com/\ldots\&lkid=\ldots & 32 & 786 \\[0.12cm]

\bottomrule
\end{tabular}
\end{table*}

\begin{table*}[ht]
\tiny
  \centering
  \caption{Codebook for the Open-Ended Response (\emph{Explain}) from the YouTube and Pinterest Experiments. \emph{Correct} Indicates Whether the Code was Marked to be Interpreted Correctly in the Participant Reponses.}
\label{tab:codebook}
\begin{tabular}{l l c}
\toprule
\textbf{Code} & \textbf{Interpretation} & \textbf{Marked \emph{Correct?}}\\
\midrule

creator\_makes\_money\_url\_purchase & The creator makes money when a product is purchased through the URL & Yes \\
creator\_makes\_money\_url\_click & The creator makes money when the URL is clicked & Yes \\
creator\_makes\_money\_sponsored & The creator makes money through sponsorships & Yes\\
creator\_makes\_money\_url\_owner & The creator makes money thorough the owner of the URL & Yes \\
creator\_makes\_money\_through\_ads & The creator makes money through ads on the URL's page & Yes\\
creator\_makes\_money\_from\_company & The creator makes money from the company selling the product & Yes \\
creator\_receives\_benefit & The creator receives ``credit'', or ``points'', or ``discounts'', or ``incentives'' & Yes \\
creator\_makes\_money\_from\_youtube & The creator makes money from YouTube for clicks on the URL & Yes \\
creator\_makes\_money\_from\_showcasing & The creator makes money from showcasing the product & Yes \\
creator\_affiliated\_company & The creator is connected or affiliated with a company (e.g., Amazon) & Yes \\
creator\_content\_sponsored & The creator's content is sponsored & Yes \\
url\_is\_advertising & The URL itself is an advertising & Yes \\
content\_paid\_ad & The content is a paid advertisement, or is paid publicity & Yes \\
third\_party\_promotion & The video is a a third-party promotion of a product & Yes \\
url\_to\_page & The URL is a URL to a product page for convenience & No \\
url\_track\_traffic & The URL is to track traffic & No \\
	
do\_not\_know & Explicit statement stating ``no idea'' or similar & No \\	
not\_clear & Respondent found the statement unclear & No \\
creator\_paying\_for\_ads & The creator is paying for ads & No \\
\bottomrule
\end{tabular}
\end{table*}

\bibliographystyle{ACM-Reference-Format}
\bibliography{sample-bibliography}

\received{April 2018} 
\received[revised]{July 2018}
\received[accepted]{September 2018} 

\end{document}